\documentclass[aps,prd,preprint,nofootinbib,11pt]{revtex4}
\usepackage{amsfonts}
\usepackage{mathrsfs}
\usepackage{graphicx}
\usepackage{amsmath}
\usepackage{amssymb}
\usepackage{subfigure}
\usepackage{epsfig}
\usepackage{graphicx}
\usepackage{color}
\newcommand{\bqa}{\begin{eqnarray}}
\newcommand{\eqa}{\end{eqnarray}}
\newcommand{\beq}{\begin{equation}}
\newcommand{\eeq}{\end{equation}}
\allowdisplaybreaks[1]
\graphicspath{{fig/}{dia/}} \DeclareGraphicsExtensions{.eps}
\begin{document}

\title{About the exotic structure of $Z_{cs}$\\[0.7cm]}

\author{Bing-Dong Wan$^1$ and Cong-Feng Qiao$^{1,2}$\footnote{qiaocf@ucas.ac.cn}\vspace{+3pt}}

\affiliation{$^1$ School of Physics, University of Chinese Academy of Science, Yuquan Road 19A, Beijing 10049 \\
$^2$ CAS Center for Excellence in Particle Physics, Beijing 10049, China}

\author{~\\~\\}

\begin{abstract}
\vspace{0.5cm}

Very recently a new hadronic structure around $3.98$ GeV was observed in BESIII experiment. From its decay modes, it is reasonable for people to assign it to the category of exotic state, say $Z^+_{cs}$, the stranged-parter of $Z_{c}(3900)$.
This finding indicates for the first time the exotic state with strange quark in charm sector, and hence has a peculiar importance. By virtue of the QCD Sum Rule technique, we analyze the $Z^+_{cs}$ about its  possible configuration and physical properties, and find it could be configured as a mixture of two types of structures, $[1_c]_{\bar{c} u}\otimes[1_c]_{\bar{s} c}$ and $[1_c]_{\bar{c} c}\otimes[1_c]_{\bar{s} u}$, or $[3_c]_{\bar{c} u}\otimes[\bar{3}_c]_{\bar{s} c}$ and $[3_c]_{\bar{c} c}\otimes[\bar{3}_c]_{\bar{s} u}$, with $J^P=1^+$. Physically, it then appears to be the emergence of a compound of four possible currents in each configuration, which tells the single current evaluation of hadron spectroscopy and their decay properties are sometimes not enough. We find in both cases the energy spectra may fit well with the experimental observation, i.e. $3.98$ GeV, within the uncertainties, while noted the former is not favored by vector meson exchange model. Various $Z^+_{cs}(3980)$ decay modes are evaluated, which are critical for pinning down its configuration and left for experimental verification. We also predict the mass of $Z^0_{cs}$, the neutral partner of $Z^+_{cs}(3980)$, and analyze its dominant decay probabilities.

\end{abstract}
 \maketitle
\newpage

The establishment of quark model (QM) in the 50s of last century is is a milestone in the
exploration of micro world \cite{GellMann:1964nj,Zweig}. The spectroscopy of conventional meson and baryon in QM are as of yet gradually confirmed in experiment and are going to be complete. Entering the new millennium, with the development of technology the so-called exotic state emerges in experiment, like $X(3872)$ \cite{Choi:2003ue}, and new ones tend to appear more frequently. Now we already have a bunch of exotic-state candidates waiting for clarification, similar to the phase of "particle zoo" in last century. To discover more exotic states and explore their properties are one of the most intriguing and important topics in particle physics for nowadays physicists, which may promote our understanding of quantum chromodynamics (QCD) and enrich our knowledge of handon spectroscopy.

In light hadron sector, because normally the spacings between various states are small, it is hard to discriminate the exotic states from the conventional hadrons, except the former possess some peculiar quantum numbers. In contrast, the exotic states in heavy hadron sector may have relatively distinct signatures. Indeed, in recent years a bunch of so-called plethora charmonium-/bottomonium-like states XYZ are observed in experiment \cite{Choi:2003ue, Aubert:2005rm, Belle:2011aa, Ablikim:2013mio, Liu:2013dau}, which provides a new horizon for our understanding of the emergence of structures in quantum chromodynamics (QCD).

Very recently, by scrutinizing the $D$-, $D_s$- and $K$-meson production in electron-position collision, BESIII Collaboration obsereved a structure in $D$ and $D_s$ invariant mass of about $3.982$ GeV with $5.3\; \sigma$ significance and decay width of some $12.8$ MeV \cite{Ablikim:2020Zcs}. If it is true, the new structure should be a charged hidden charm state, and hence named as $Z^+_{cs}(3980)$. From its known decay products the new state most likely possesses a quantum number of $1^+$. If the BESIII observation is further confirmed to be a hadronic structure, rather the kinematic effect, it turns out to be a remarkable discovery in the exploration of hadron spectroscopy, the novel strange-hidden-charm state.

About strange-hidden-charm(bottom) states there have been some investigations in the literature \cite{Lee:2008uy,Dias:2013qga,Ferretti:2020ewe,Cao:2017lui,Voloshin:2019ilw,Di:2019qwv,SanchezSanchez:2017xtl,Tang:2019nwv,Chen:2013wca}. Nevertheless, before the experimental evidence appearing, theoretical investigations tend to be with large uncertainties, say diverse yields in different theoretical frameworks. The BESIII observation in some sense rules out the tetraquark octet-octet configuration \cite{Tang:2019nwv}, at least the $Z^+_{cs}(3980)$ has a weak coupling to that kind of current. Considering the previous calculation results deviate more or less from the experimental measurement, we find the new structure could be a compound of states from four molecular currents, which may then be evaluated by virtue of the model independent Shifman, Vainshtein and Zakharov (SVZ) QCD sum rule technique \cite{Shifman}.

The SVZ sum rule, viz QCD sum rule, has some peculiar advantages in exploring hadron properties involving nonpertubative QCD. It is a QCD based theoretical framework which incorporates nonperturbative effects universally order by order, rather a phenomenological model, and has already achieved a lot in the study of hadron spectroscopy and decays. To establish the sum rules, the starting point is to construct the proper interpolating currents corresponding to the hadron of interest. Using the current, one can then construct the two-point correlation function, which has two representations: the QCD representation and the phenomenological representation. Then, roughly speaking, by equating these two representations, the QCD sum rules will be formally established, from which the hadron mass and decay width may be deduced.

In this work, we make a thorough analysis on the $Z^+_{cs}(3980)$ in the framework of QCD sum rule, including mass spectroscopy and decay properties. Composite currents in molecular and tetraquark configurations are taken into account. Its neutral partner $Z^0_{cs}$ is also evaluated for future confirmation. In order to analyze the mass spectrum of $Z^+_{cs}$ state, one has to construct the appropriate current for it. The lowest order possible interpolating currents of $1^+$ charged charmonium-like strange molecular state take the following four forms:
\begin{eqnarray}
j_\mu^{\bar{D}^{\ast 0} D_s^+} &=& i[\bar{c}_a \gamma_\mu u_a][\bar{s}_b \gamma_5 c_b]\, , \label{current-A1} \\
j_\mu^{\bar{D}^0 D_s^{\ast +}} &=& i[\bar{c}_a \gamma_5 u_a][\bar{s}_b \gamma_\mu c_b]\, , \label{current-B1} \\
j_\mu^{J/\psi K^+} &=& i[\bar{c}_a \gamma_\mu c_a][\bar{s}_b \gamma_5 u_b]\, , \label{current-C1} \\
j_\mu^{\eta_c K^{+\ast}} &=& i[\bar{c}_a \gamma_5 c_a][\bar{s}_b \gamma_\mu u_b]\, ,\label{current-D1}
\end{eqnarray}
where $a$  and $b$ are color indices, $\mu$ denotes Lorentz index.  Therefore, the interpolating current for $[1_c]\otimes[1_c]$ state of $Z^+_{cs}$ can be expressed as mixing of the currents in Eqs.(\ref{current-A1})-(\ref{current-D1}), i.e.,
\begin{eqnarray}
j_{\mu}^{\mathcal M} &=& \mathcal{A}_1 j_\mu^{\bar{D}^{\ast 0} D_s^+} + \mathcal{B}_1 j_\mu^{\bar{D}^0 D_s^{\ast +}} +\mathcal{C}_1 j_\mu^{J/\psi K^+} +\mathcal{D}_1 j_\mu^{\eta_c K^{+\ast}}. \label{current-mixing1}
\end{eqnarray}
On the other hand, the possible tetraquark interpolating currents can be constructed as
\begin{eqnarray}
j_\mu^{A} &=& i\epsilon_{abc}\epsilon_{dec}[u^T_a C \gamma_5 c_b][\bar{s}_d \gamma_\mu C \bar{c}_e^T]\, , \label{current-A2} \\
j_\mu^{B} &=& i\epsilon_{abc}\epsilon_{dec}[u^T_a C \gamma_\mu c_b][\bar{s}_d \gamma_5 C \bar{c}_e^T]\, , \label{current-B2} \\
j_\mu^{C} &=& i\epsilon_{abc}\epsilon_{dec}[u^T_a C  c_b][\bar{s}_d \gamma_\mu \gamma_5 C \bar{c}_e^T]\, , \label{current-C2} \\
j_\mu^{D} &=& i\epsilon_{abc}\epsilon_{dec}[u^T_a C \gamma_\mu \gamma_5 c_b][\bar{s}_d C \bar{c}_e^T]\, .\label{current-D2}
\end{eqnarray}
Here, the superscripts $A$, $B$, $C$, and $D$ indicate the currents composed of $0^+\otimes 1^+$, $1^+\otimes 0^+$, $0^-\otimes 1^-$, and $1^-\otimes 0^-$, respectively, and $C$ represents the charge conjugation matrix. Therefore, the interpolating current for $[3_c]\otimes[\bar{3}_c]$ state of $Z^+_{cs}$ can be expressed as mixing of the currents in Eqs.(\ref{current-A2})-(\ref{current-D2}), i.e.,
\begin{eqnarray}
j_{\mu}^{\mathcal T} &=& \mathcal{A}_2 j_\mu^{A} + \mathcal{B}_2 j_\mu^{B} +\mathcal{C}_2 j_\mu^{C} +\mathcal{D}_2 j_\mu^{D}. \label{current-mixing2}
\end{eqnarray}
Here, the superscript ${\mathcal M}$ and  ${\mathcal T}$ denote the molecular and tetraquark state, respectively. It is noteworthy that in the literature most of the calculations were performed by single-current analysis. However, in fact, different inner configurations may yield different physical results, not to say their interference. Before the advent of experimental measurement, to make a comprehensive analysis on a typical hadron is usually unrealistic, but now for $Z^+_{cs}$ we can do so, which is important in order to know its real structure.

With the currents of (\ref{current-mixing1}) and (\ref{current-mixing2}), the two-point correlation function can be readily established, i.e.,
\begin{eqnarray}
\Pi_{\mu \nu}(q) &=& i \int d^4 x e^{i q \cdot x} \langle 0 | T {\{} j_{\mu}(x), j_{\nu}(0)^\dagger {\}} |0 \rangle \nonumber \\
&=& (q_\mu q_\nu -q^2 g_{\mu \nu}) \Pi(q^2)\; ,
\end{eqnarray}
where $|0\rangle$ denotes the physical vacuum. In the partonic representation, the dispersion relation may express the correlation function $\Pi(q^2)$ as
\begin{eqnarray}
\Pi_i^{OPE} (q^2) &=& \int_{s_{min}}^{\infty} d s
\frac{\rho_i^{OPE}(s)}{s - q^2} + \Pi_i^{sum}(q^2)\; .
\label{OPE-hadron}
\end{eqnarray}
Here, $\rho_i^{OPE}(s) = \text{Im} [\Pi_i^{OPE}(s)] / \pi$ and $\Pi_i^{sum }(q^2)$ is the sum of those contributions in the correlation function that have no imaginary part but have nontrivial magnitudes after the Borel transformation. $s_{min}$ is a kinematic limit, which usually corresponds to the square of the sum of the current quark masses of the hadron \cite{Albuquerque:2013ija}, i.e., $s_{min}=(2 m_c + m_s + m_u)^2$. In (\ref{OPE-hadron}) the subscript $i$ represents ${\mathcal M}$ and ${\mathcal T}$ for molecular and tetraquark states, respectively. By applying the Borel transformation to (\ref{OPE-hadron}), we then have
\begin{eqnarray}
\Pi_i^{OPE}(M_B^2)\!\! = \!\!\int_{s_{min}}^{\infty} d s
\rho_i^{OPE}(s)e^{- s / M_B^2} + \Pi_i^{sum }(M_B^2)\ . \label{quark-gluon}
\end{eqnarray}

In the hadronic representation, after isolating the ground state contribution from the hadronic state, we obtain the correlation function $\Pi(q^2)$ in dispersion integral over the physical region, i.e.,
\begin{eqnarray}
\Pi_i(q^2) & = & \frac{\lambda_i^2}{M_{i}^2 - q^2} + \frac{1}{\pi} \int_{s_0}^\infty d s \frac{\rho_i(s)}{s - q^2} \; , \label{hadron}
\end{eqnarray}
where  $M^{i}$ denotes the mass the lowest lying hadronic state, either molecule-like or tetraquark state, $\rho_i(s)$ is the spectral density that contains the contributions from higher excited states and the continuum states above the threshold $s_0$. The coupling constant $\lambda_i$ is defined through $\langle 0 | j_\mu^{i} | Z^+_{cs} \rangle = \lambda^{i} \epsilon_\mu$.

By performing the Borel transform on the hadronic side, Eq.(\ref{hadron}), and matching it to Eq.(\ref{quark-gluon}), we can then obtain the mass and the coupling constant of the tetraqark state,
\begin{eqnarray}
M_i(s_0, M_B^2) &=& \sqrt{- \frac{L_{i1}(s_0, M_B^2)}{L_{i0}(s_0, M_B^2)}} \; ,\\ \label{mass-Eq}
\lambda_{i}^2  e^{-M_{i}^2 / M_B^2} &=& L_{i0}(s_0, M_B^2) \; , \label{lambda-Eq}
\end{eqnarray}
where the moments $L_1$ and $L_0$ are,  respectively, defined as:
\begin{eqnarray}
L_{i0}(s_0, M_B^2) & = & \int_{s_{min}}^{s_0} d s \; \rho_i^{OPE}(s) e^{-
s / M_B^2} + \Pi_i^{sum}(M_B^2) \; , \label{L0}
\\ L_{i1}(s_0, M_B^2) & = & \frac{\partial}{\partial{\frac{1}{M_B^2}}}{L_{i0}(s_0, M_B^2)} \; .
\end{eqnarray}

In the numerical calculation of QCD sum rules, the values of input parameters we take are \cite{Shifman, Reinders:1984sr, P.Col, Narison:1989aq}: $m_u = 2.3 \; \text{MeV}$, $m_d = 6.4 \; \text{MeV}$, $m_s(2 \, \text{GeV}) = (95 \pm 5) \; \text{MeV}$, $m_c (m_c) = \overline{m}_c= (1.275 \pm 0.025)
\; \text{GeV}$, $\langle \bar{q} q \rangle = - (0.24 \pm 0.01)^3
\; \text{GeV}^3$, $\langle \bar{s} s \rangle = (0.8 \pm 0.1)
\langle \bar{q} q \rangle$, $\langle g_s^2 G^2 \rangle = 0.88
\; \text{GeV}^4$, $\langle \bar{q} g_s \sigma \cdot G q
\rangle = m_0^2 \langle \bar{q} q \rangle$, $\langle \bar{s} g_s \sigma \cdot G s
\rangle = m_0^2 \langle \bar{s} s \rangle$, $\langle g_s^3 G^3
\rangle = 0.045 \; \text{GeV}^6$, and $m_0^2 = 0.8 \; \text{GeV}^2$,
in which the $\overline{\text{MS}}$ running heavy quark masse is adopted.

Moreover, there exist two additional parameters $M_B^2$ and $s_0$ introduced in establishing the sum rules, which will be fixed in light of the so-called standard procedures abiding by two criteria \cite{Shifman, Reinders:1984sr, P.Col,Albuquerque:2013ija}. The first one asks for the convergence of the OPE. That is, one needs to compare individual contributions with the total magnitude on the OPE side, and choose a reliable region for $M_B^2$ to retain the convergence. The second criterion requires that the portion of lowest lying pole contribution (PC), the ground state contribution, in the total, pole plus continuum, should be over 50\%~\cite{P.Col, Matheus:2006xi,Albuquerque:2013ija}. The two criteria can be formulated as
\begin{eqnarray}
  R_i^{OPE} = \left| \frac{L_{i0}^{dim=8}(s_0, M_B^2)}{L_{i0}(s_0, M_B^2)}\right|\, ,
\end{eqnarray}
\begin{eqnarray}
  R_i^{PC} = \frac{L_{i0}(s_0, M_B^2)}{L_{i0}(\infty, M_B^2)} \; . \label{RatioPC}
\end{eqnarray}

Various $s_0$ satisfying above constraints should be taken into account in the numerical analysis. 
Among these values, we need to pick up the one which yields an optimal window for Borel parameter
$M^2_B$. That is to say, in the optimal window, the tetraquark mass $M_Z$ is somehow
independent of the Borel parameter $M^2_B$ 
In practice, we may vary $\sqrt{s_0}$ by $0.1$ GeV in numerical calculation \cite{Wan:2019ake,Tang:2019nwv,Qiao:2013dda,Qiao:2013raa,Tang:2016pcf}, which sets the upper and lower 
bounds and hence the uncertainties of $\sqrt{s_0}$.

\begin{figure}
\includegraphics[width=6.8cm]{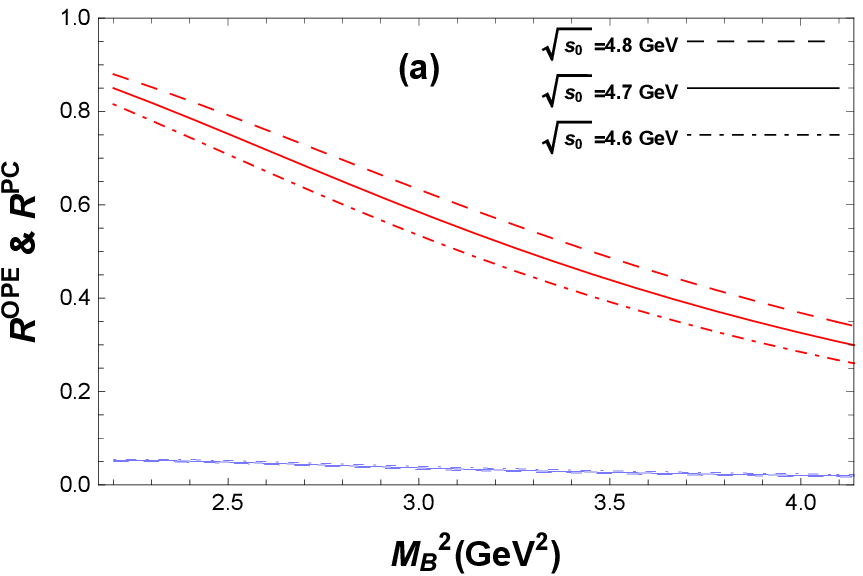}
\includegraphics[width=6.8cm]{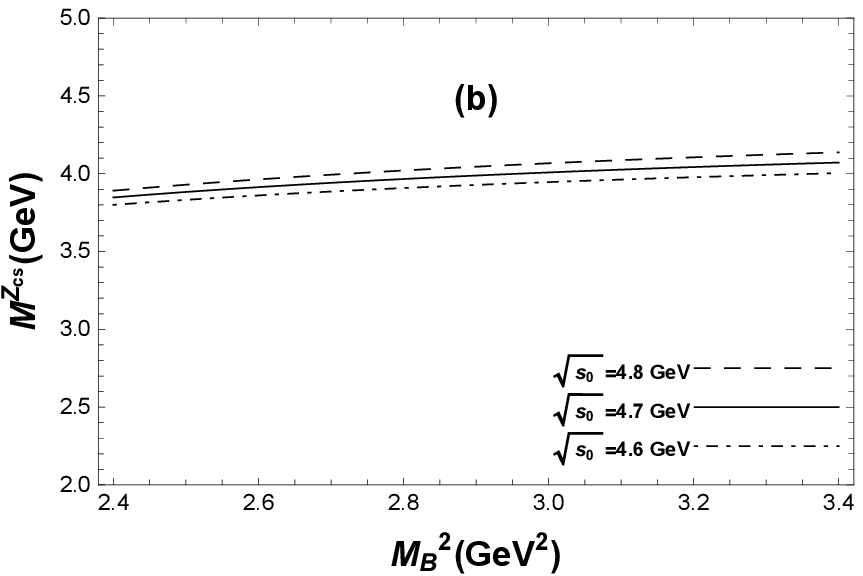}
\caption{ (a) The ratios ${R_{\mathcal M}^{OPE}}$ and ${R_{\mathcal M}^{PC}}$ as functions of the Borel parameter $M_B^2$ for different values of $\sqrt{s_0}$, where blue lines represent ${R_{\mathcal M}^{OPE}}$ and red lines denote ${R_{\mathcal M}^{PC}}$ . (b) The mass of $Z^+_{cs}$ as a function of the Borel parameter $M_B^2$ for different values of $\sqrt{s_0}$.} \label{fig1}
\end{figure}

\begin{figure}
\includegraphics[width=6.8cm]{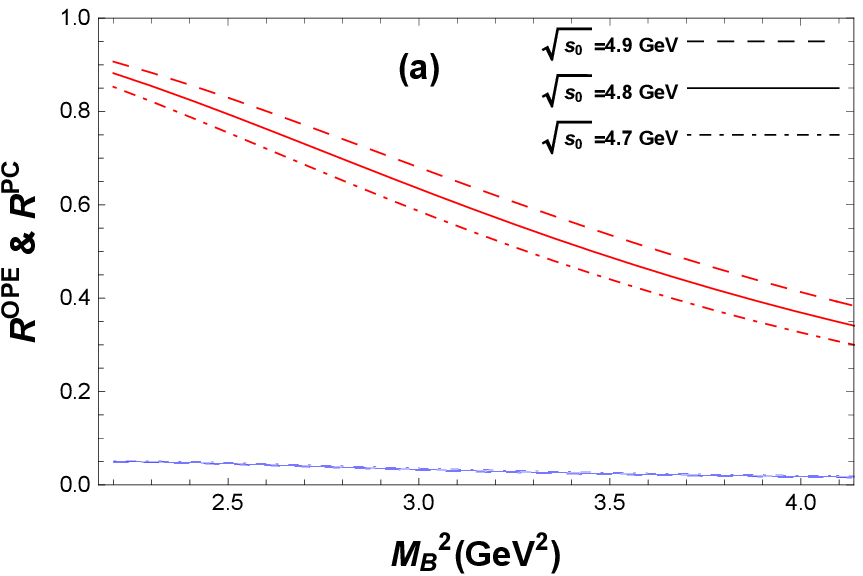}
\includegraphics[width=6.8cm]{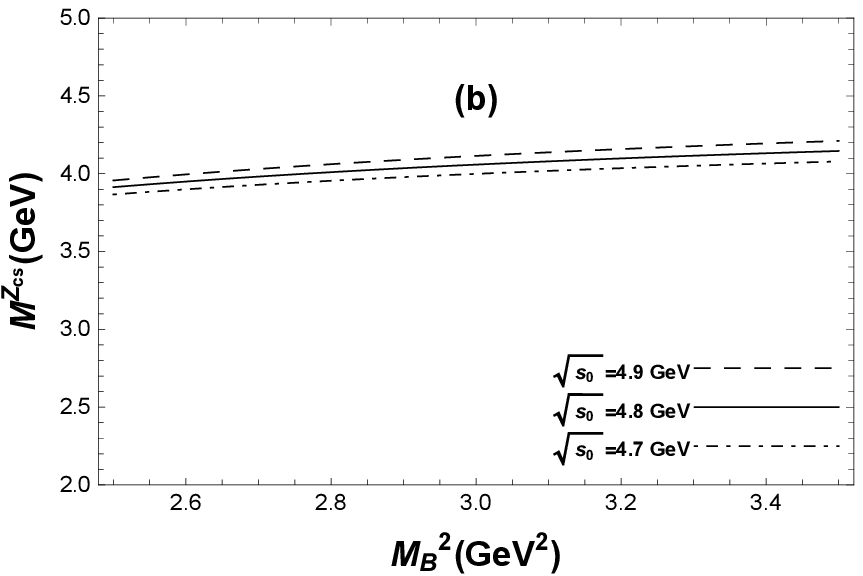}
\caption{The same caption as in Fig \ref{fig1}, but for the neutral $Z^0_{cs}$.} \label{fig2}
\end{figure}

With above preparation we numerically evaluate the mass spectrum of the $Z^+_{cs}$ with different $(\mathcal{A}_1, \mathcal{B}_1, \mathcal{C}_1, \mathcal{D}_1)$ and $(\mathcal{A}_2, \mathcal{B}_2, \mathcal{C}_2, \mathcal{D}_2)$ for molecular and tetraquark states, respectively. The ratios $R_{\mathcal M}^{OPE}$ and $R_{\mathcal M}^{PC}$ of the molecular state $Z^+_{cs}$  are shown as functions of Borel parameter $M_B^2$ in Fig. \ref{fig1}(a)  with $\mathcal{A}_1=0.47$, $\mathcal{B}_1=-0.47$, $\mathcal{C}_1=0.53$, and $\mathcal{D}_1=-0.53$ and with different values of $\sqrt{s_0}$, $4.6$, $4.7$, and $4.8$ GeV. The dependency relations between $Z^+_{cs}$ mass and parameter $M_B^2$ are given in Fig. \ref{fig1}(b). The optimal window for   Borel parameter was found at $2.5 \le M_B^2 \le 3.3\; \rm{GeV}^2$, and the mass and coupling constant of $Z^+_{cs}$ are extracted as
\begin{eqnarray}
&&M_{\mathcal M}^{Z^+_{cs}} = (3.98 \pm 0.14) \, \text{GeV}  \; , \\  \label{eq-mass-1}
&&\lambda_{\mathcal M}^{Z^+_{cs}} = (1.81 \pm 0.12)\times 10^{-2} \, \rm{GeV}^5 \; .
\end{eqnarray}
Replace $u-$quark with $d-$quark in Eq.(\ref{current-mixing1}), its neutral partner $Z^0_{cs}$ will be obtained. With the same mixing coefficients, the ratios $R_{\mathcal M}^{OPE}$ and $R_{\mathcal M}^{PC}$ of $Z^0_{cs}$  are shown as functions of Borel parameter $M_B^2$ in Fig. \ref{fig2}(a) with different values of $\sqrt{s_0}$, $4.7$, $4.8$, and $4.9$ GeV and the dependency relations between $Z^0_{cs}$ mass and parameter $M_B^2$ are given in Fig. \ref{fig2}(b). The optimal window for   Borel parameter was found at $2.6 \le M_B^2 \le 3.4\; \rm{GeV}^2$, and the mass and coupling constant of $Z^0_{cs}$ are extracted as
\begin{eqnarray}
&&M_{\mathcal M}^{Z^0_{cs}} = (4.03 \pm 0.16) \, \text{GeV}  \; , \\
&&\lambda_{\mathcal M}^{Z^0_{cs}} = (1.95 \pm 0.11)\times 10^{-2} \, \rm{GeV}^5 \; .
\end{eqnarray}

\begin{figure}
\includegraphics[width=6.8cm]{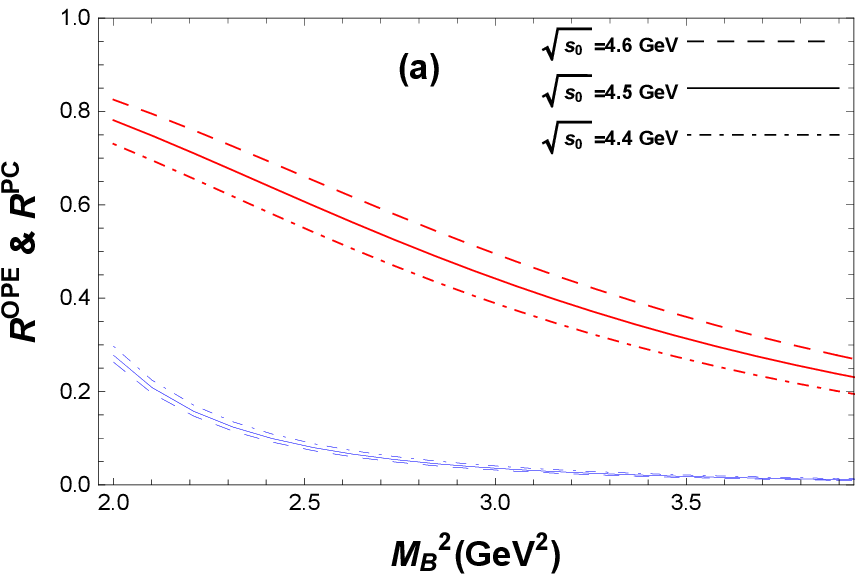}
\includegraphics[width=6.8cm]{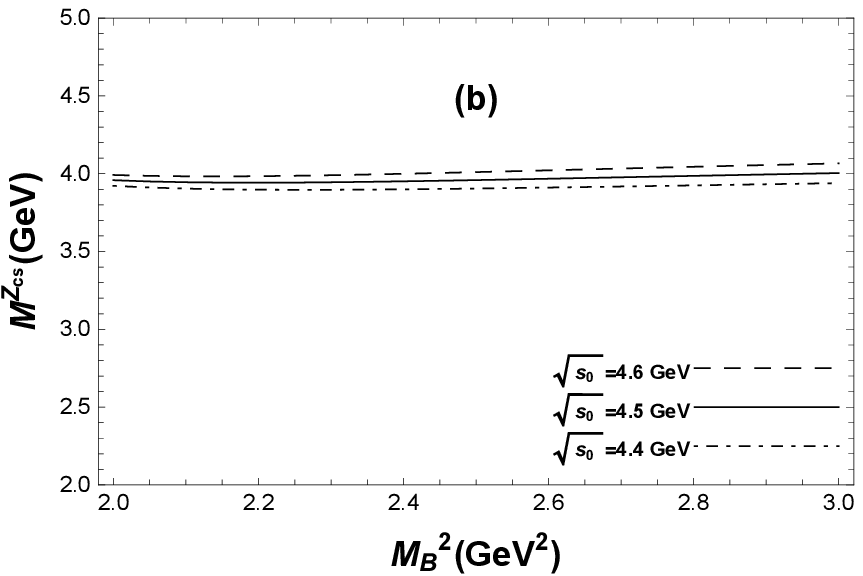}
\caption{ The same caption as in Fig \ref{fig1}, but in tetraquark structure.} \label{fig3}
\end{figure}

\begin{figure}
\includegraphics[width=6.8cm]{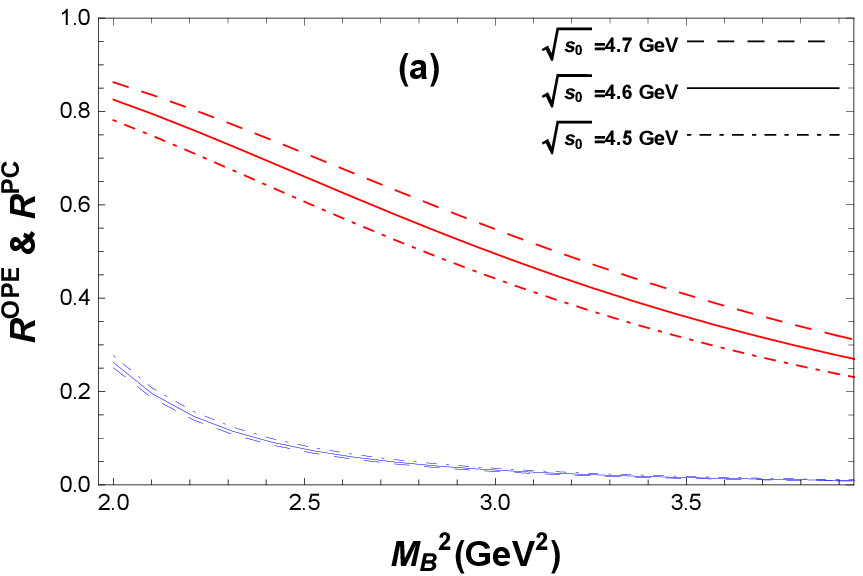}
\includegraphics[width=6.8cm]{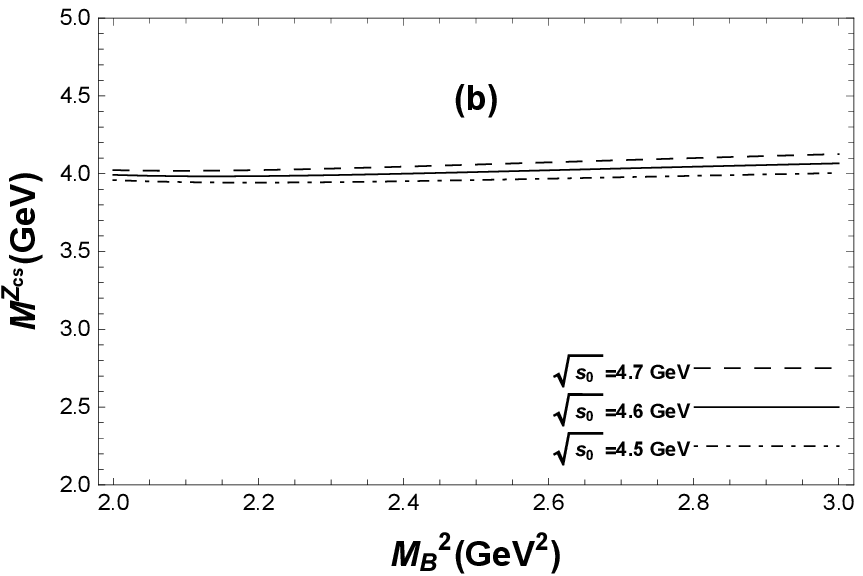}
\caption{ The same caption as in Fig \ref{fig2}, but for neutral $Z^0_{cs}$ in tetraquark structure.} \label{fig4}
\end{figure}

On the other hand, the ratios $R_{\mathcal T}^{OPE}$ and $R_{\mathcal T}^{PC}$ of the compact tetraquark, the diquark-antidiquark, state $Z^+_{cs}$  are shown as functions of Borel parameter $M_B^2$ in Fig. \ref{fig3}(a) with $\mathcal{A}_2=0.61$, $\mathcal{B}_2=-0.61$, $\mathcal{C}_2=-0.36$, and $\mathcal{D}_2=0.36$ and with different values of $\sqrt{s_0}$, $4.4$, $4.5$, and $4.6$ GeV. The dependency relations between $Z^+_{cs}$ mass and parameter $M_B^2$ are given in Fig. \ref{fig3}(b). The optimal window for Borel parameter is found at $2.1 \le M_B^2 \le 3.0\; \rm{GeV}^2$, and the mass and coupling constant of $Z^+_{cs}$ are extracted as
\begin{eqnarray}
&&M_{\mathcal T}^{Z^+_{cs}}=  (3.98 \pm 0.08) \, \text{GeV}  \; , \\  \label{eq-mass-2}
&&\lambda_{\mathcal T}^{Z^+_{cs}} = (1.29 \pm 0.10)\times 10^{-2} \, \rm{GeV}^5 \; .
\end{eqnarray}
Its neutral partner $Z^0_{cs}$ will be obtained by replacing $u-$quark with $d-$quark in Eq.(\ref{current-mixing2}). With the same mixing coefficients, the ratios $R_{\mathcal T}^{OPE}$ and $R_{\mathcal T}^{PC}$ of $Z^0_{cs}$ are shown as functions of Borel parameter $M_B^2$ in Fig. \ref{fig4}(a) with different values of $\sqrt{s_0}$, $4.5$, $4.6$, and $4.7$ GeV and the dependencies between $Z^0_{cs}$ mass and parameter $M_B^2$ are given in Fig. \ref{fig4}(b). The optimal window for Borel parameter is found at $2.1 \le M_B^2 \le 3.0\; \rm{GeV}^2$, and the mass and coupling constant of $Z^0_{cs}$ are therefore obtained as
\begin{eqnarray}
&&M_{\mathcal T}^{Z^0_{cs}}= (4.04 \pm 0.09) \, \text{GeV}  \; , \\
&&\lambda_{\mathcal T}^{Z^0_{cs}} = (1.43 \pm 0.10)\times 10^{-2} \, \rm{GeV}^5 \; .
\label{tetraneu}
\end{eqnarray}
In above results, (\ref{eq-mass-1})-(\ref{tetraneu}), errors stem from the uncertainties of the quark masses, the condensates and the threshold parameter $\sqrt{s_0}$.

Note, as shown in Appendices, that the cross terms $\mathcal{A}_1 \mathcal{B}_1$($\mathcal{A}_2 \mathcal{C}_2$) and $\mathcal{C}_1 \mathcal{D}_1$($\mathcal{B}_2 \mathcal{D}_2$) give no contribution to the mass spectrum, while $\mathcal{A}_1 \mathcal{C}_1$, $\mathcal{A}_1 \mathcal{D}_1$, $\mathcal{B}_1 \mathcal{C}_1$($\mathcal{A}_2 \mathcal{B}_2$, $\mathcal{A}_2 \mathcal{D}_2$, $\mathcal{B}_2 \mathcal{C}_2$), and $\mathcal{B}_1 \mathcal{D}_1$($\mathcal{C}_2 \mathcal{D}_2$) terms do.

The calculation of the decay vertex starts from the three-point correlation function,
\begin{eqnarray}
  \Pi^i_{\mu \nu} (q, q_1, q_2) = \int d^4x d^4y\;  e^{iq_1 \cdot x+i q_2 \cdot y} \Pi^i_{\mu \nu}(x,y)\ , \label{three-point}
\end{eqnarray}
where $\Pi^i_{\mu \nu}(x,y)= \langle 0|T[j_\mu^{\bar{D}^\ast}(x) j_5^{D_s^+}(y) j_\nu^{i }(0)^\dagger]|0\rangle$ and $q = q_1 + q_2$. The interpolating currents of $\bar{D}^\ast$ and $D_s^+$ take the following forms:
\begin{eqnarray}
  j_\mu^{\bar{D}^\ast} &=& \bar{c}_a \gamma_\mu u_a \, , \\
  j_5^{D_s^+} &=& i \bar{s}_b \gamma_5 c_b \ .
\end{eqnarray}

On the phenomenological side of the QCD sum rules, we insert the intermediate states into the correlation function (\ref{three-point}), and obtain
\begin{eqnarray}
      \Pi_{\mu \nu}^{phen} &=& \frac{ -\lambda^{Z^+_{cs}} m_{D^\ast} f_{D^\ast} f_{D_s} m_{D_s}^2 g_{Z^+_{cs} \bar{D}^\ast D_s}}{(m_c +m_s) (q^2 - M_{Z_{cs}}^{2})(q_1^2 - m_{D^\ast}^2) (q_2^2 - m_{D_s}^2)} \nonumber \\
      &\times& \bigg(- g_{\mu \alpha} + \frac{q_{1\mu} q_{1\alpha}}{m_{D^\ast}^2}\bigg)\bigg(-g_{\nu}^{\alpha} + \frac{q_{\nu} q^{\alpha}}{M_{Z_{cs}}^{2}}\bigg) \ . \label{three-point-p}
\end{eqnarray}
Here, $g_{Z^+_{cs} \bar{D}^\ast D_s}$ presents $Z_{cs}$ decay form factor; $f_{D^\ast}$ and $f_{D_s}$ are meson decay constants defined as:
\begin{eqnarray}
      \langle \bar{D}^\ast D_s | Z^+_{cs} \rangle &=& g_{Z^+_{cs} \bar{D}^\ast D_s} \varepsilon^\ast_\alpha(q_1) \varepsilon^\alpha(q) \; ,\\
      \langle 0 | j_\mu^{\bar{D}^\ast} | \bar{D}^\ast \rangle &=& m_{D^\ast} f_{D^\ast} \varepsilon_\mu(q_1) \; ,\\
      \langle 0 | j_\mu^{D_s} | D_s \rangle &=& \frac{f_{D_s} m_{D_s}^2}{m_c+m_s} \; .
\end{eqnarray}
On the OPE side of QCD sum rules, the three-point function can be formulated as \cite{Bracco:2011pg}
\begin{eqnarray}
      \Pi^{OPE} = \int d v d s \frac{\rho^{OPE}(s,v)}{(s-q_1^2)(v-q_2^2)}\ .  \label{three-point-o}
\end{eqnarray}

Performing Borel transforms $q_1^2=q_2^2\to M_B^2$  on both side of (\ref{three-point-p}) and (\ref{three-point-o}), we obtain the form factor $g_{Z^+_{cs} \bar{D}^\ast D_s}$ by equating $\Pi^{OPE}(s_0,M_B^2)$ to $\Pi^{phen}(s_0,M_B^2)$, where $s_0$ is the continuum threshold and $M_B^2$ is the Borel parameter of the $D$ meson. Details of the decay widths calculation are presented in the Appendix for reference.

In numerical evaluation of the $Z_{cs}$ decays, we adopt the mass and decay constants employed in Refs.\cite{Novikov:1977dq,Deshpande:1994mk,Bordes:2005wi,Blossier:2009bx,pdg}, i.e., $m_{D^\ast} = 2.01 \, \text{GeV}$, $ m_{D_s} = 1.97 \, \text{GeV}$, $m_{D^\ast_s} = 2.11 \, \text{GeV}$, $m_{D} = 1.86 \, \text{GeV}$, $ m_{J/\psi} = 3.07 \, \text{GeV}$, $m_{\eta_c} = 2.98 \, \text{GeV}$, $m_{K^\ast} = 0.89 \, \text{GeV}$, $m_{K} = 0.49 \, \text{GeV}$, $ f_{D^\ast} = 0.24 \, \text{GeV}$, $f_{D_s} = 0.24 \, \text{GeV}$, $f_{D^\ast_s} = 0.33 \, \text{GeV}$, $f_{D} = 0.18 \, \text{GeV}$, $f_{D^\ast} = 0.41 \, \text{GeV}$, $f_{D_s} = 0.35 \, \text{GeV}$, $f_{K^\ast} = 0.22 \, \text{GeV}$, and $f_{K} = 0.16 \, \text{GeV}$.

With the above inputs and formula in Appendix we can readily obtain the various form factors and decay widths, as presented in Table \ref{tab1}. It is worth mentioning that the molecular $Z_{cs}$ decay widths and form factors are give for the first time in this work. The tetraquark $Z_{cs}$ decays were once evaluated by Dias, {\it et al}. \cite{Dias:2013qga}, however since they adopted only two currents, that is only (\ref{current-A2}) and (\ref{current-B2}), the $Z_{cs}$ to $J/\psi K^+$ and $\eta_c K^\ast$ decay widths are quite different. Furthermore, we find there is a misprint in the contribution of mixed condensate in ref.\cite{Dias:2013qga}, the gluon in light quark radiation is missed, whatsoever its numerical effects are tiny.
\begin{table}
\begin{center}
\caption{Form factors and the decay widths of $Z_{cs}$.}
\label{tab1}
\begin{tabular}{|c|c|c|c|c|c|c}\hline\hline
                                                         Decay mode & Form factor $g$ (\rm{GeV})   & Decay width $\Gamma$ (\rm{MeV})  \\ \hline
  $Z^{\mathcal M}_{cs} \to \bar{D}^\ast D_s$   &              $ 3.72\pm 0.78$          &  $3.23 \pm 1.29 $\\ \hline
  $Z^{\mathcal M}_{cs} \to \bar{D} D_s^\ast$   &              $-4.18\pm 0.88$          &  $4.08 \pm 1.65 $\\ \hline
  $Z^{\mathcal M}_{cs} \to  J/\psi K^+$             &              $ 1.68\pm 0.39$          &  $4.92 \pm 2.23 $\\ \hline
  $Z^{\mathcal M}_{cs} \to \eta_c K^\ast$          &              $-2.84\pm 0.69$          &  $8.42 \pm 3.84 $\\ \hline
  $Z^{\mathcal T}_{cs} \to \bar{D}^\ast D_s$   &              $ 2.09\pm 0.42 $          &  $1.14 \pm 0.44 $\\ \hline
  $Z^{\mathcal T}_{cs} \to \bar{D} D_s^\ast$    &              $ 1.30 \pm 0.30$          &  $0.40 \pm0.18 $\\ \hline
  $Z^{\mathcal T}_{cs} \to J/\psi K^+$              &              $ 0.74 \pm 0.22$          &  $0.98 \pm 0.54 $\\ \hline
  $Z^{\mathcal T}_{cs} \to \eta_c K^\ast$          &              $ 3.83 \pm 1.07$          &  $15.60 \pm 8.08 $\\ \hline
 \hline
\end{tabular}
\end{center}
\end{table}

In summary, we perform a complete analysis on the $Z^+_{cs}(3980)$ in the framework of QCD sum rule, which matches well the recent experimental observation in BESIII experiment. In our calculation, the full leading order currents in various configurations are taken into account, their relative weights are estimated by fitting the QCD sum rule results to the experimental measurements. We consider both molecular and tetraquark structures with different configurations, that is set $Z^+_{cs}(3980)$ in $[1_c]_{\bar{c} u}\otimes[1_c]_{\bar{s} c}$ and $[1_c]_{\bar{c} c}\otimes[1_c]_{\bar{s} u}$, or $[3_c]_{\bar{c} u}\otimes[\bar{3}_c]_{\bar{s} c}$ and $[3_c]_{\bar{c} c}\otimes[\bar{3}_c]_{\bar{s} u}$ representation with $J^P=1^+$.  Within the error of uncertainties, the summed width of four dominant decay channels falls in the experimental measurement, that is $\Gamma_{\mathcal M}=(20.65 \pm 9.01)$ MeV and $\Gamma_{\mathcal T}=(18.12 \pm 9.24)$ MeV for molecular and tetraquark states, respectively. The mass spectrum of the $Z^+_{cs}$'s neutral partner $Z^0_{cs}$ is also calculated. Note, with the results of this work, future experimental measurements on $Z^+_{cs}$ dominant decay channels may pin down its inner configurations.

\vspace{0.7cm} {\bf Acknowledgments}

This work was supported in part by the National Natural Science Foundation of China(NSFC) under the Grants 11975236 and 11635009.

\newpage
\begin{widetext}
\appendix
\section{The spectral densities of $Z^+_{cs}$}
In order to calculate the spectral density of the operator product expansion (OPE) side, the light quark ($q=u$, $d$ or $s$) and heavy-quark ($Q=c$ or $b$) full propagators $S^q_{i j}(x)$ and $S^Q_{i j}(p)$ are employed, say
\begin{eqnarray}
S^q_{j k}(x) \! \! & = & \! \! \frac{i \delta_{j k} x\!\!\!\slash}{2 \pi^2
x^4} - \frac{\delta_{jk} m_q}{4 \pi^2 x^2} - \frac{i t^a_{j k} G^a_{\alpha\beta}
}{32 \pi^2 x^2}(\sigma^{\alpha \beta} x\!\!\!\slash
+ x\!\!\!\slash \sigma^{\alpha \beta}) - \frac{\delta_{jk}}{12} \langle\bar{q} q \rangle + \frac{i\delta_{j k}
x\!\!\!\slash}{48} m_q \langle \bar{q}q \rangle - \frac{\delta_{j k} x^2}{192} \langle g_s \bar{q} \sigma \cdot G q \rangle \nonumber \\ &+& \frac{i \delta_{jk} x^2 x\!\!\!\slash}{1152} m_q \langle g_s \bar{q} \sigma \cdot G q \rangle - \frac{t^a_{j k} \sigma_{\alpha \beta}}{192}
\langle g_s \bar{q} \sigma \cdot G q \rangle
+ \frac{i t^a_{jk}}{768} (\sigma_{\alpha \beta} x \!\!\!\slash + x\!\!\!\slash \sigma_{\alpha \beta}) m_q \langle
g_s \bar{q} \sigma \cdot G q \rangle \;,
\end{eqnarray}
\begin{eqnarray}
S^Q_{j k}(p) \! \! & = & \! \! \frac{i \delta_{j k}(p\!\!\!\slash + m_Q)}{p^2 - m_Q^2} - \frac{i}{4} \frac{t^a_{j k} G^a_{\alpha\beta} }{(p^2 - m_Q^2)^2} [\sigma^{\alpha \beta}
(p\!\!\!\slash + m_Q)
+ (p\!\!\!\slash + m_Q) \sigma^{\alpha \beta}] \nonumber \\ &+& \frac{i\delta_{jk}m_Q  \langle g_s^2 G^2\rangle}{12(p^2 - m_Q^2)^3}\bigg[ 1 + \frac{m_Q (p\!\!\!\slash + m_Q)}{p^2 - m_Q^2} \bigg] \nonumber \\ &+& \frac{i \delta_{j k}}{48} \bigg\{ \frac{(p\!\!\!\slash +
m_Q) [p\!\!\!\slash (p^2 - 3 m_Q^2) + 2 m_Q (2 p^2 - m_Q^2)] }{(p^2 - m_Q^2)^6}
\times (p\!\!\!\slash + m_Q)\bigg\} \langle g_s^3 G^3 \rangle \; .
\end{eqnarray}
Here, the vacuum condensates are explicitly shown. For more explanation on above propagators, readers may refer to Refs.~\cite{Albuquerque:2013ija}.

\begin{eqnarray}
\rho^{OPE}(s) & = & \rho^{pert}(s) + \rho^{\langle \bar{q} q
\rangle}(s) + \rho^{\langle \bar{s} s
\rangle}(s) + \rho^{\langle G^2 \rangle}(s) + \rho^{\langle \bar{q} G q \rangle}(s) \nonumber \\
&+& \rho^{\langle \bar{s} G s \rangle}(s)
+ \rho^{\langle \bar{q} q \rangle \langle \bar{s} s \rangle}(s) + \rho^{\langle G^3 \rangle}(s)+ \rho^{\langle \bar{q} q\rangle \langle G^2 \rangle}(s) + \rho^{\langle \bar{s} s\rangle \langle G^2 \rangle}(s) \nonumber\\
&+& \rho^{\langle \bar{s} s \rangle \langle \bar{q} G q \rangle}(s)+ \rho^{\langle \bar{q} q \rangle \langle \bar{s} G s \rangle}(s)\; , \label{rho-OPE} \\
\Pi^{sum}(q^2) &=&  \Pi^{\langle \bar{q} q \rangle \langle \bar{s} s \rangle}(q^2)+ \Pi^{\langle G^3 \rangle}(q^2) +\Pi^{\langle \bar{q} q\rangle \langle G^2 \rangle}(q^2)+\Pi^{\langle \bar{s} s\rangle \langle G^2 \rangle}(q^2)\nonumber\\
&+& \Pi^{\langle \bar{q} q \rangle \langle \bar{s} G s \rangle}(q^2) + \Pi^{\langle \bar{s} s \rangle \langle \bar{q} G q \rangle}(q^2)  \; . \label{rho-OPE-Pi}
\end{eqnarray}

\subsection{The spectral densities for molecular state}

The spectral density $\rho^{OPE}(s)$ is calculated up to dimension eight.

\begin{eqnarray}
 \rho^{pert}(s) &=& \frac{1}{3 \times 2^{13} \pi^6}\int^{\alpha_{max}}_{\alpha_{min}} \frac{d \alpha}{\alpha^3} \int^{1 - \alpha}_{\beta_{min}} \frac{d \beta}{\beta^3} {\cal F}^3_{\alpha \beta} (\alpha + \beta - 1) \bigg{\{} 18  {\cal A}^2 \bigg[{\cal F}_{\alpha \beta}(1+\alpha+\beta)- 4m_c [2 \alpha m_q\nonumber\\
  &+&\beta m_s (1+\alpha+\beta)]\bigg] + 18  {\cal B}^2 \bigg[{\cal F}_{\alpha \beta}(1+\alpha+\beta) -4m_c [2 \beta m_s +\alpha m_q (1+\alpha+\beta)]\bigg] \nonumber\\
 &+& 6 \bigg[ 3({\cal C}^2+{\cal D}^2){\cal F}_{\alpha \beta}(1+\alpha+\beta) +4m_c^2(\alpha + \beta - 1)[3{\cal C}^2+{\cal D}^2(2+\alpha+\beta)] \bigg]\nonumber\\
 &+&{\cal AC}\bigg[3{\cal F}_{\alpha \beta}(1+\alpha+\beta)+12 m_c\big[m_c(\alpha + \beta - 1)-(m_q+m_s)(\beta^2+\beta+\alpha\beta+2\alpha)\big]\bigg]\nonumber\\
 &+&{\cal BC}\bigg[3{\cal F}_{\alpha \beta}(1+\alpha+\beta)+12 m_c\big[m_c(\alpha + \beta - 1)-(m_q+m_s)(\alpha^2+\alpha+\alpha\beta+2\beta)\big]\bigg]\nonumber\\
  &+&{\cal AD}\bigg[3{\cal F}_{\alpha \beta}(1+\alpha+\beta)+4 m_c\big[m_c(\alpha + \beta - 1)(\alpha + \beta + 2)\nonumber\\
  &-&3(\alpha+\beta)[2m_q+m_s(1+\alpha+\beta)]\big]\bigg]+{\cal BD}\bigg[3{\cal F}_{\alpha \beta}(1+\alpha+\beta)\nonumber\\
  &+&4 m_c\big[m_c(\alpha + \beta - 1)(\alpha + \beta + 2)-3(\alpha+\beta)[2m_s+m_q(1+\alpha+\beta)]\big]\bigg]\bigg{\}}\; ,\\
 \rho^{\langle \bar{q} q \rangle}(s) &=& \frac{\langle \bar{q} q \rangle}{2^9\pi^4} \int^{\alpha_{max}}_{\alpha_{min}} d \alpha \bigg{\{} \int^{1 - \alpha}_{\beta_{min}} d \beta \bigg{\{} \frac{6{\cal A}^2{\cal F}_{\alpha \beta}(2m_c {\cal F}_{\alpha \beta}+\beta m_q {\cal F}_{\alpha \beta}-4\beta m_c^2 m_s )}{\alpha\beta^2}\nonumber\\
 &+&\frac{6{\cal B}^2{\cal F}_{\alpha \beta}(2m_c (\alpha+\beta) {\cal F}_{\alpha \beta}+\beta m_q {\cal F}_{\alpha \beta}-4\beta m_c^2 m_s )}{\alpha\beta^2}+\frac{6{\cal D}^2{\cal F}_{\alpha \beta}({\cal F}_{\alpha \beta}+2m_c^2)(m_q-2m_s)}{\alpha\beta}\nonumber\\
 &+&\frac{6{\cal C}^2{\cal F}_{\alpha \beta}(m_q {\cal F}_{\alpha \beta}+2 m_q m_c^2(\alpha+\beta)-4 m_c^2 m_s )}{\alpha\beta}\nonumber\\
 &+&\frac{{\cal AC}{\cal F}_{\alpha \beta}[{\cal F}_{\alpha \beta}(\alpha\beta m_q +2m_c(\alpha+\beta))+2\alpha\beta m_c^2(m_q(\alpha+\beta)-2m_s)]}{\alpha^2\beta^2}\nonumber\\
 &+&\frac{{\cal AD}{\cal F}_{\alpha \beta}[\alpha\beta({\cal F}_{\alpha \beta} + 2m_c^2)(m_q-2m_s) +2m_c{\cal F}_{\alpha \beta}(\alpha+\alpha\beta+\beta^2)]}{\alpha^2\beta^2}\nonumber\\
 &+&\frac{{\cal BC}{\cal F}_{\alpha \beta}[{\cal F}_{\alpha \beta}(\alpha\beta m_q +2m_c(\alpha+\beta)^2)+2\alpha\beta m_c^2(m_q(\alpha+\beta)-2m_s)]}{\alpha^2\beta^2}\nonumber\\
  &+&\frac{{\cal BD}{\cal F}_{\alpha \beta}[\alpha\beta({\cal F}_{\alpha \beta} + 2m_c^2)(m_q-2m_s) +2m_c{\cal F}_{\alpha \beta}(\alpha^2+\alpha\beta+\beta)]}{\alpha^2\beta^2}\bigg{\}}\nonumber\\
  &+&\frac{{\cal H}_\alpha^2}{\alpha(\alpha-1)}\bigg[6({\cal A}^2+{\cal B}^2+{\cal C}^2+{\cal D}^2)m_q-12{\cal C}^2m_s\nonumber\\
  &+&({\cal AC+AD+BC+BD})(m_q-2m_s)\bigg]\bigg{\}}\;,\\
 \rho^{\langle \bar{s} s \rangle}(s) &=& \frac{\langle \bar{s} s \rangle}{2^9\pi^4} \int^{\alpha_{max}}_{\alpha_{min}} d \alpha \bigg{\{} \int^{1 - \alpha}_{\beta_{min}} d \beta \bigg{\{} \frac{ -24 {\cal A}^2 m_c^2 m_q{\cal F}_{\alpha \beta}}{\alpha\beta}-\frac{6{\cal D}^2{\cal F}_{\alpha\beta}({\cal F}_{\alpha\beta}+2 m_c^2)(2m_q-m_s)}{\alpha\beta}\nonumber\\
 &+&\frac{6{\cal C}^2{\cal F}_{\alpha\beta}(m_s {\cal F}_{\alpha\beta}-4m_c^2m_q+2m_c^2m_s(\alpha+\beta))}{\alpha\beta}\nonumber\\
 &+&\frac{6{\cal B}^2{\cal F}_{\alpha\beta}(2m_c {\cal F}_{\alpha\beta}-4\alpha m_c^2m_q+\alpha m_s {\cal F}_{\alpha\beta})}{\alpha^2\beta}\nonumber\\
  &+&\frac{{\cal AC}{\cal F}_{\alpha \beta}[{\cal F}_{\alpha \beta}(\alpha\beta m_s +2m_c(\alpha+\beta)^2)+2\alpha\beta m_c^2(m_s(\alpha+\beta)-2m_q)]}{\alpha^2\beta^2}\nonumber\\
 &+&\frac{{\cal AD}{\cal F}_{\alpha \beta}[\alpha\beta({\cal F}_{\alpha \beta} + 2m_c^2)(m_s-2m_q) +2m_c{\cal F}_{\alpha \beta}(\alpha+\alpha\beta+\beta^2)]}{\alpha^2\beta^2}\nonumber\\
 &+&\frac{{\cal BC}{\cal F}_{\alpha \beta}[{\cal F}_{\alpha \beta}(\alpha\beta m_s +2m_c(\alpha+\beta))+2\alpha\beta m_c^2(m_s(\alpha+\beta)-2m_q)]}{\alpha^2\beta^2}\nonumber\\
  &+&\frac{{\cal BD}{\cal F}_{\alpha \beta}[\alpha\beta({\cal F}_{\alpha \beta} + 2m_c^2)(m_s-2m_q) +2m_c{\cal F}_{\alpha \beta}(\alpha^2+\alpha\beta+\beta)]}{\alpha^2\beta^2}\bigg{\}}\nonumber\\
   &+&\frac{{\cal H}_\alpha^2}{\alpha(\alpha-1)}\bigg[6({\cal B}^2+{\cal C}^2-{\cal D}^2)m_s-12({\cal C}^2+{\cal D}^2)m_q\nonumber\\
&+&({\cal AC+AD+BC+BD})(m_s-2m_q)\bigg]\bigg{\}}\;,\\
\rho^{\langle G^2 \rangle}(s) &=& \frac{\langle g_s^2 G^2\rangle}{3^2\times2^{14}\pi^6} \int^{\alpha_{max}}_{\alpha_{min}} d \alpha \bigg{\{} \int^{1 - \alpha}_{\beta_{min}} \frac{d \beta}{\alpha^3\beta^3} \bigg{\{} 36{\cal A}^2\bigg[\alpha\beta{\cal F}_{\alpha\beta}\big[2\alpha\beta m_c m_s (\alpha+\beta-2)\nonumber\\
&-&{\cal F}_{\alpha\beta}(\alpha^2-2\alpha+4\alpha\beta+3\beta^2)\big]-m_c(\alpha+\beta-1)(\alpha+\beta+1)\big[ 3\beta^3m_s{\cal F}_{\alpha\beta}\nonumber\\
&-&m_c({\cal F}_{\alpha\beta}+\beta m_c m_s)(\alpha^3+\beta^3)  \big]\bigg]+ 36{\cal B}^2\bigg[\alpha\beta{\cal F}_{\alpha\beta}[2\beta{\cal F}_{\alpha\beta}+6\alpha\beta m_c m_s\nonumber\\
&-&(\alpha+\beta)(3\alpha+\beta){\cal F}_{\alpha\beta}]+(\alpha+\beta-1)m_c[-6\beta^3 m_s {\cal F}_{\alpha\beta}+(\alpha+\beta)(1+\alpha+\beta)\nonumber\\
&\times&(\alpha^2-\alpha\beta+\beta^2)m_c{\cal F}_{\alpha\beta}-2 \beta m_c^2 m_s (\alpha^3+\beta^3) ]\bigg]+{\cal C}^2 \bigg[  12m_c^4 (\alpha+\beta-1)^2\nonumber\\
&\times&(\alpha+\beta)(\alpha+\beta+2)(\alpha^2-\alpha\beta+\beta^2)+18\alpha\beta{\cal F}^2_{\alpha\beta}(3\alpha^2+3\beta^2+8\alpha\beta-3)\nonumber\\
&+&36m_c^2{\cal F}_{\alpha\beta}[2\alpha^5+7\alpha^4\beta-4\alpha^3+13\alpha^3\beta^2+2\beta^2-4\beta^3+2\beta^5\nonumber\\
&+&\alpha^2 \beta ((\beta - 1) (13\beta + 9) + 2) + \alpha\beta (7\beta^3 - 9\beta + 4)] \bigg] +{\cal D}^2\bigg[18\alpha\beta{\cal F}_{\alpha\beta}({\cal F}_{\alpha\beta} (8 \alpha  \beta\nonumber\\
&+&(\alpha -4) \alpha +\beta ^2-4 \beta +3)+12 \alpha  \beta  m_c^2)+ 36 m_c^2 (\alpha +\beta -1) [{\cal F}_{\alpha\beta} (\alpha ^4\nonumber\\
&+&\alpha ^3 (\beta +4)+3 \alpha ^2 (\beta -1)+\alpha  \beta ^2 (\beta +3)+\beta ^2 (\beta  (\beta +4)-3))\nonumber\\
&+&m_c^2 (\alpha +\beta -1) (\alpha +\beta )(\alpha ^2-\alpha  \beta +\beta ^2)] \bigg]-{\cal AC}\bigg[6m_c(\alpha+\beta-1)[3m_s{\cal F}_{\alpha\beta}(2 \alpha ^3\nonumber\\
&+&(\alpha +1) \beta ^3+\beta ^4)-m_c^3(\alpha^3+\beta^3)(m_c(\alpha+\beta-1)-m_s(\beta^2+\beta+2\alpha\nonumber\\
&+&\alpha\beta))-{\cal F}_{\alpha\beta}m_c(\alpha ^4+\alpha ^3 (\beta +4)
+3 \alpha ^2 (\beta -1)+\alpha  \beta ^2 (\beta +3)\nonumber\\
&+&\beta ^2 (\beta  (\beta +4)-3))]+3\alpha\beta{\cal F}_{\alpha\beta}[{\cal F}_{\alpha\beta}(5 \alpha ^2+\alpha  (8 \beta -4)+\beta  (5 \beta -4)+3)\nonumber\\
&+&2m_c(2m_c(2 (\alpha -1) \beta ^2+2 (\alpha -1) \alpha  \beta +\alpha  ((\alpha -2) \alpha +3)+\beta ^3+3 \beta -2)\nonumber\\
&+&m_s(\alpha+\beta)(\alpha ^2+2 \alpha  (\beta +2)+\beta  (\beta +4)-3))]\bigg]+{\cal AD}\bigg[-3\alpha\beta{\cal F}^2_{\alpha\beta}[3 \alpha ^2\nonumber\\
&+&\alpha  (8 \beta -4)+\beta  (11 \beta +4)-3]+6 m_c^3 (\alpha +\beta -1) (\alpha +\beta ) (\alpha ^2\nonumber\\
&-&\alpha  \beta +\beta ^2) [m_c (\alpha +\beta -1)-m_s (\alpha  (\beta +2)
+\beta ^2+\beta)]+6m_c {\cal F}^2_{\alpha\beta}[m_c (\alpha ^5\nonumber\\
&+&\alpha ^4 (6 \beta +3)+\alpha ^3 (\beta +1) (9 \beta -7)+\alpha ^2 (\beta +1) (\beta  (5 \beta -9)+3)\nonumber\\
&+&2 \alpha  \beta ^2 ((\beta -3) \beta +3)+(\beta -1) \beta ^2 (\beta  (\beta +4)-3))-m_s (\alpha +\beta ) (2 \alpha ^3 (\beta +3)\nonumber\\
&-&3 \alpha ^2 \left(\beta ^2+2\right)
-2 \alpha  (\beta -1) \beta  (\beta +3)+3 \beta ^2 (\beta ^2-1))]\bigg]+{\cal BC}\bigg[2 m_c^3 (\alpha +\beta\nonumber\\
 &-&1) (\alpha +\beta ) (\alpha ^2-\alpha  \beta +\beta ^2) [m_c (\alpha+\beta -1) (\alpha +\beta +2)-6 m_s (\alpha +\beta )]\nonumber\\
 &-&3\alpha\beta{\cal F}^2_{\alpha\beta}[5 \alpha ^2+\alpha  (8 \beta -4)+\beta  (5 \beta -4)+3]
+6m_c{\cal F}_{\alpha\beta}[m_c (2 \alpha ^2 (\alpha ^3-2 \alpha +1)\nonumber\\
&+&3 \alpha  \beta ^4+(\alpha  (\alpha +4)-4) \beta ^3+(\alpha  (\alpha  (\alpha +4)-9)+2) \beta ^2\nonumber\\
&+&\alpha  (\alpha  (\alpha  (3 \alpha +4)-9)+4) \beta +2 \beta ^5)-6 m_s (\alpha +\beta ) ((\alpha -1) \alpha ^2+(\beta -1) \beta ^2)]\bigg]\nonumber\\
&+&{\cal BD}\bigg[-3 \alpha\beta{\cal F}_{\alpha\beta}[11 \alpha ^2+\alpha  (8 \beta +4)+\beta  (3 \beta -4)-3]\nonumber\\
&+&6m_c{\cal F}_{\alpha\beta}[m_c (2 \alpha ^2 ((\alpha -3) \alpha +3) \beta
+(\alpha -1) \alpha ^2 (\alpha  (\alpha +4)-3)\nonumber\\
&+&3 (2 \alpha +1) \beta ^4+(\alpha +1) (9 \alpha -7) \beta ^3+(\alpha +1) (\alpha  (5 \alpha -9)+3) \beta ^2\nonumber\\
&+&\beta ^5)-m_s (\alpha +\beta ) (-3 (\alpha ^2+2) \beta ^2+3 \alpha ^2 (\alpha ^2-1)+2 (\alpha +3) \beta ^3\nonumber\\
&-&2 (\alpha -1) \alpha  (\alpha +3) \beta )]
+6m_c^3(\alpha+\beta-1)(\alpha+\beta)(\alpha ^2-\alpha  \beta +\beta ^2)[m_c (\alpha\nonumber\\
&+&\beta -1)-m_s (\alpha  (\alpha +\beta +1)+2 \beta )]\bigg]\bigg{\}}\nonumber\\
&+&\frac{6{\cal H}^2_\alpha(-{\cal C} ({\cal A} + {\cal B} + 6 {\cal C}) + 3 ({\cal A} + {\cal B}) {\cal D} + 18 {\cal D}^2)}{(\alpha -1) \alpha}\bigg{\}}\;,\\
\rho^{\langle \bar{q} G q \rangle}(s) &=& \frac{\langle g_s \bar{q} \sigma \cdot G q \rangle}{3\times2^9\pi^4} \int_{\alpha_{min}}^{\alpha_{max}} \bigg{\{} \int_{\beta_{min}}^{1 - \alpha} d \beta \bigg{\{}\frac{18 {\cal B}^2 m_c (2\beta m_c m_s -{\cal F}_{\alpha\beta}(2\alpha+3\beta))}{\beta^2} \nonumber\\
&-& 6{\cal D}^2 m_c^2 m_q +{\cal AC}\frac{2 {\cal F}_{\alpha\beta}(\alpha+\beta)(m_c(\alpha+2\beta-1)-\alpha m_s)-\alpha^2\beta m_c^2 m_q}{\alpha^2\beta}\nonumber\\
&+&{\cal AD}\frac{-{\cal F}_{\alpha\beta} m_c (\alpha -5 \beta +1)+{\cal F}_{\alpha\beta} m_s (\alpha +3 \beta )-2 \beta  m_c^3 (\alpha +\beta )+6 \beta  m_c^2 m_s}{\alpha\beta}\nonumber\\
&-&{\cal BC}\frac{m_c \left({\cal F}_{\alpha\beta} \left(6 \alpha ^2+11 \alpha  \beta +2 \beta ^2\right)+2 \beta  m_c (\alpha +\beta ) (\beta  m_c-3 m_s)\right)}{\alpha\beta^2}\nonumber\\
&+&\frac{{\cal BD}}{\alpha^2\beta^2}[{\cal F}_{\alpha\beta} \left(\alpha  \beta  m_s (3 \alpha +\beta )-m_c \left(6 \alpha ^3+2 (1-3 \alpha ) \beta ^2+(4 \alpha +1) \alpha  \beta -4 \beta ^3\right)\right)\nonumber\\
&-&2 \alpha  \beta  m_c^2 (\beta  m_c (\alpha +\beta )-3 \alpha  m_s)]\bigg{\}}+6{\cal C}^2[2m_c^2(m_q-3m_s)+{\cal H}_\alpha(9m_s-2m_q)]\nonumber\\
&-&12{\cal D}^2({\cal H}_\alpha-m_c^2)(m_q-3m_s)+6({\cal A}^2+{\cal B}^2)[\frac{3m_c {\cal H}_\alpha}{1-\alpha}- 2 m_q {\cal H}_\alpha+m_c^2(m_q-3m_s)]\nonumber\\
&+&{\cal AC}[2 m_c^2 (m_q-3m_s)+{\cal H}_\alpha(9 m_s-2 m_q)-\frac{{\cal H}_\alpha(2m_c-m_s)}{\alpha(\alpha-1)}]\nonumber\\
&+&{\cal AD}[2 m_c^2 (m_q-3m_s)+\frac{{\cal H}_\alpha(3m_s+2m_c(2+\alpha)+2\alpha(m_s+m_q(\alpha-1)-3\alpha m_s))}{\alpha(\alpha-1)}]\nonumber\\
&+&{\cal BC}\frac{-2 \alpha  {\cal H}_\alpha (m_c+(\alpha -1) m_q)+(6 (\alpha -1) \alpha -3) {\cal H}_\alpha m_s+2 (\alpha -1) \alpha m_c^2 (m_q-3 m_s)}{\alpha(\alpha-1)}\nonumber\\
&+&\frac{{\cal BD}}{\alpha(\alpha-1)}[{\cal H}_\alpha (2 (\alpha -3) m_c+2 \alpha  (-\alpha  m_q+m_q+(3 \alpha -5) m_s)+m_s)\nonumber\\
&+&2 (\alpha -1) \alpha  m_c^2 (m_q-3 m_s)]\bigg{\}}\;,\\
\rho^{\langle \bar{s} G s \rangle}(s) &=& \frac{\langle g_s \bar{s} \sigma \cdot G s \rangle}{3^2\times2^9\pi^4} \int_{\alpha_{min}}^{\alpha_{max}} \bigg{\{} \int_{\beta_{min}}^{1 - \alpha} d \beta \bigg{\{} -\frac{48{\cal F}_{\alpha\beta}m_c(3\alpha+2\beta)}{\alpha^2}\nonumber\\
&-&3{\cal AC}\frac{(\alpha+\beta)(3m_c^2m_s\alpha (\alpha+\beta) + {\cal F}_{\alpha\beta}(8\alpha m_c+ 3\alpha m_s+ 6\beta m_c))}{\alpha^2\beta}\nonumber\\
&+& \frac{3{\cal AD}}{\alpha^2\beta^2}  [-\alpha  \beta  m_c^2 \left(2 \beta  m_c (\alpha +\beta )+m_s \left(-2 \alpha ^2-2 \alpha  \beta +\alpha +3 \beta \right)\right)\nonumber\\
&+&{\cal F}_{\alpha\beta}(m_c \left(4 \alpha ^3+\alpha ^2 (3 \beta -2)-\alpha  \beta  (\beta +1)-6 \beta ^3\right)-\alpha  \beta  m_s (\alpha +3 \beta ))]\nonumber\\
&+&\frac{3{\cal BC}}{\alpha\beta^2} [{\cal F}_{\alpha\beta}(m_c \left(4 \alpha ^2+\alpha  (6 \beta -2)+\beta  (5 \beta -2)\right)-\beta  m_s (\alpha +\beta ))\nonumber\\
&-&m_c^2\beta(2 \beta  m_c (\alpha +\beta )+m_s \left((\alpha -2) \beta +(\alpha -2) \alpha +\beta ^2\right))]\nonumber\\
&+&\frac{3{\cal BD}}{\alpha\beta}[{\cal F}_{\alpha\beta}(5 \alpha  m_c-\beta  (m_c+m_s)-\text{mc}-3 \alpha  m_s)\nonumber\\
&-&m_c^2(2 \alpha  m_c (\alpha +\beta )+m_s \left(-2 \alpha  \beta +3 \alpha -2 \beta ^2+\beta \right))] \bigg{\}}\nonumber\\
&+&48({\cal A}^2+{\cal B}^2) \frac{m_c({\cal H}_\alpha-\alpha m_c m_q)}{\alpha} + 48{\cal C}^2m_q(3{\cal H}_\alpha-2m_c^2)\nonumber\\
&+&96{\cal D}^2 m_q({\cal H}_\alpha-m_c^2) +{\cal AC}[-2 m_c^2 (9 m_q+m_s) +3{\cal H}_\alpha(9  m_q\nonumber\\
&+&2m_s  +\frac{3 m_s}{\alpha(\alpha-1)}) ]+3{\cal AD}[-2 m_c^2 (3 m_q-4m_s) +{\cal H}_\alpha(6m_q)\nonumber\\
&+&\frac{m_s(3+4\alpha-8\alpha^2)-2m_c(\alpha+2)}{\alpha(\alpha-1)}] -3{\cal BC}[6m_c^2m_q\nonumber\\
&+&\frac{{\cal H}_\alpha[m_s+2(m_c+3m_q+m_s)\alpha+2\alpha^2(3m_q+m_s)]}{\alpha(\alpha-1)}]+3{\cal BD}[-2m_c^2(3m_q-4m_s)\nonumber\\
&+&\frac{{\cal H}_\alpha(2 (\alpha -3)m_c+6 (\alpha -1) \alpha  m_q+(4 (3-2 \alpha ) \alpha -1) m_s)}{\alpha(\alpha-1)}]\bigg{\}}\;,\\
\rho^{\langle \bar{q} q \rangle \langle \bar{s} s \rangle}(s) &=& \frac{\langle \bar{q} q \rangle \langle \bar{s} s \rangle}{3\times2^6\pi^2} \int_{\alpha_{min}}^{\alpha_{max}} d \alpha \bigg{\{}6{\cal A}^2 m_c(-2m_c+m_q-\alpha m_q+2\alpha m_s)\nonumber\\
&+&6{\cal B}^2 m_c(-2m_c+2m_q-2\alpha m_q+\alpha m_s)+{\cal C}^2(36{\cal H}_\alpha-24m_c^2)+{\cal D}^2(24{\cal H}_\alpha-24m_c^2)\nonumber\\
&+&{\cal AC}(6 {\cal H}_\alpha+m_c (-4 m_c+m_q+2m_s))+{\cal AD}(4 {\cal H}_\alpha+m_c (-4 m_c+m_q(1+\alpha)\nonumber\\
&+&m_s(1+\alpha)))+{\cal BC}(6 {\cal H}_\alpha+m_c (-4 m_c+2m_q+m_s))\nonumber\\
&+&{\cal BD}(4 {\cal H}_\alpha+m_c (-4 m_c+m_q(2-\alpha)+m_s(2-\alpha)))\bigg{\}}\;,\\
\rho^{\langle G^3 \rangle}(s) &=& \frac{\langle g_s^3 G^3 \rangle}{3\times2^{15}\pi^6} \int_{\alpha_{min}}^{\alpha_{max}} \frac{d\alpha}{\alpha^3} \int_{\beta_{min}}^{1 - \alpha} \frac{d \beta}{\beta^3} \bigg{\{} 6{\cal A}^2 (\alpha +\beta -1) (\alpha +\beta +1)\nonumber\\
&\times& [{\cal F}_{\alpha\beta} \left(\alpha ^3+\beta ^3\right)+m_c \left(2 m_c \left(\alpha ^4+\beta ^4\right)-\beta  m_s \left(\alpha ^3+6 \beta ^3\right)\right)]\nonumber\\
&+& 6{\cal B}^2 (\alpha +\beta -1) [{\cal F}_{\alpha\beta} (\alpha +\beta ) (\alpha +\beta +1) \left(\alpha ^2-\alpha  \beta +\beta ^2\right)\nonumber\\
&+&2 m_c \left(m_c (\alpha +\beta +1) \left(\alpha ^4+\beta ^4\right)-\beta  m_s \left(\alpha ^3+6 \beta ^3\right)\right)]\nonumber\\
&+&6{\cal C}^2 (\alpha +\beta -1)[{\cal F}_{\alpha\beta} (\alpha +\beta ) (\alpha +\beta +1) \left(\alpha ^2-\alpha  \beta +\beta ^2\right)\nonumber\\
&+&2 m_c^2 (2 \alpha ^5+\alpha ^4 (3 \beta +2)+\alpha ^3 \left(\beta ^2+\beta -2\right)+\alpha ^2 \beta ^3\nonumber\\
&+&\alpha  \beta ^3 (3 \beta +1)+2 \beta ^3 (\beta ^2+\beta -1))]+6{\cal D}^2 (\alpha +\beta -1)[{\cal F}_{\alpha\beta} (\alpha +\beta ) (\alpha +\beta +1)\nonumber\\
&\times& \left(\alpha ^2-\alpha  \beta +\beta ^2\right)+2 m_c^2 (\alpha ^5+\alpha ^4 (\beta +4)+3 \alpha ^3 (\beta -1)+\alpha  \beta ^3 (\beta +3)\nonumber\\
&+&\beta ^3 (\beta  (\beta +4)-3))]+{\cal AC}(\alpha +\beta -1)[{\cal F}_{\alpha\beta} (\alpha +\beta ) (\alpha +\beta +1) \left(\alpha ^2-\alpha  \beta +\beta ^2\right)\nonumber\\
&+&m_c (2 m_c (\alpha ^3 \beta ^2+\alpha ^3 (3 \alpha +1) \beta +\left(\alpha ^2+\alpha -2\right) \beta ^3+2 \alpha ^3 \left(\alpha ^2+\alpha -1\right)\nonumber\\
&+&(3 \alpha +2) \beta ^4+2 \beta ^5)-m_s (\alpha +\beta +1) \left(2 \alpha ^2+3 \alpha  \beta +2 \beta ^2\right) \left(3 \alpha ^2-4 \alpha  \beta +3 \beta ^2\right))]\nonumber\\
&+&{\cal AD}(\alpha +\beta -1)[{\cal F}_{\alpha\beta} (\alpha +\beta ) (\alpha +\beta +1) \left(\alpha ^2-\alpha  \beta +\beta ^2\right)\nonumber\\
&+&m_c (2m_c \left(\alpha ^5+\alpha ^4 (\beta +4)+3 \alpha ^3 (\beta -1)+\alpha  \beta ^3 (\beta +3)+\beta ^3 (\beta  (\beta +4)-3)\right)\nonumber\\
&-&m_s \left(\alpha ^4 (\beta +12)+\alpha ^3 \beta  (\beta +1)+2 \alpha  \beta ^3 (3 \beta +1)+6 \beta ^4 (\beta +1)\right))]\nonumber\\
&+&{\cal BC}(\alpha +\beta -1)[{\cal F}_{\alpha\beta} (\alpha +\beta ) (\alpha +\beta +1) \left(\alpha ^2-\alpha  \beta +\beta ^2\right)\nonumber\\
&+&2 m_c (m_c (\alpha ^3 \beta ^2+\alpha ^3 (3 \alpha +1) \beta +\left(\alpha ^2+\alpha -2\right) \beta ^3+2 \alpha ^3 \left(\alpha ^2+\alpha -1\right)\nonumber\\
&+&(3 \alpha +2) \beta ^4+2 \beta ^5)-m_s \left(6 \alpha ^4+\alpha ^3 \beta +\alpha  \beta ^3+6 \beta ^4\right))]\nonumber\\
&+&{\cal BD}(\alpha +\beta -1)[{\cal F}_{\alpha\beta} (\alpha +\beta ) (\alpha +\beta +1) \left(\alpha ^2-\alpha  \beta +\beta ^2\right)\nonumber\\
&+&m_c (2 m_c \left(\alpha ^5+\alpha ^4 (\beta +4)+3 \alpha ^3 (\beta -1)+\alpha  \beta ^3 (\beta +3)+\beta ^3 (\beta  (\beta +4)-3)\right)\nonumber\\
&-&m_s \left(6 (\alpha +1) \alpha ^4+2 (3 \alpha +1) \alpha ^3 \beta +(\alpha +12) \beta ^4+(\alpha +1) \alpha  \beta ^3\right))]\bigg{\}}\;,\\
\rho^{\langle \bar{q} q\rangle \langle G^2 \rangle}(s) &=& \frac{\langle \bar{q} q\rangle \langle G^2 \rangle}{3^2\times2^{10}\pi^4}  \int_{\alpha_{min}}^{\alpha_{max}} d\alpha \bigg{\{}\int_{\beta_{min}}^{1 - \alpha} d \beta \bigg{\{}-18{\cal A}^2\frac{\alpha m_c}{\beta^2} +6{\cal B}^2\frac{m_c(3\alpha^2+3\alpha\beta-\beta^2)}{\beta^2}\nonumber\\
&-&6{\cal D}^2m_s+3{\cal AC}\frac{m_c(\alpha+\beta)(\alpha^2+\beta^2)}{\alpha^2\beta^2}+{\cal AD}[\frac{m_c(\alpha+\beta)(3\alpha^2+3\beta^2-2\alpha\beta(1+\beta))}{\alpha^2\beta^2}\nonumber\\
&-&m_s]+{\cal BC}\frac{m_c(\alpha+\beta)^2(3\alpha^2-\alpha\beta+3\beta^2)}{\alpha^2\beta^2}+{\cal BD}[-m_s\nonumber\\
&+&\frac{m_c(\alpha+\beta)(3\alpha^2+3\beta^2-2\alpha\beta(1+\alpha))}{\alpha^2\beta^2}] \bigg{\}} +(18{\cal A}^2-6{\cal B}^2) m_c  +(18{\cal C}^2-6{\cal D}^2) m_s\nonumber\\
&+&3{\cal AC}[m_s+\frac{m_c}{\alpha(1-\alpha)}]+{\cal AD}[-m_s+\frac{m_c(3-4\alpha)}{\alpha(1-\alpha)}]+{\cal BC}[3m_s+\frac{-m_c}{\alpha(1-\alpha)}]\nonumber\\
&+&{\cal BD}[-m_s+\frac{m_c(4\alpha-1)}{\alpha(1-\alpha)}]\bigg{\}}\;,\\
\rho^{\langle \bar{s} s\rangle \langle G^2 \rangle}(s) &=& \frac{\langle \bar{s} s\rangle \langle G^2 \rangle}{3^2\times2^{11}\pi^4}  \int_{\alpha_{min}}^{\alpha_{max}} d\alpha \bigg{\{}\int_{\beta_{min}}^{1 - \alpha} d \beta \bigg{\{} 12{\cal A}^2 \frac{m_c(3\beta^2+3\alpha\beta-\alpha^2)}{\alpha^2} + 36{\cal B}^2\frac{m_c \beta}{\alpha^2}\nonumber\\
&+&18{\cal C}^2m_s+6{\cal D}^2m_s +{\cal AC}[3ms + \frac{2m_c(\alpha+\beta)^2(3\alpha^2-\alpha\beta+3\beta^2)}{\alpha^2\beta^2}] \nonumber\\
&+&{\cal AD}[ms+\frac{2m_c(\alpha+\beta)(3\alpha^2+3\beta^2-2\alpha\beta(1+\beta))}{\alpha^2\beta^2}]\nonumber\\
&+&{\cal BC}[3ms + \frac{6m_c(\alpha+\beta)(\alpha^2+3\beta)}{\alpha^2\beta^2}]+{\cal BD}[ms\nonumber\\
&+&\frac{2m_c(\alpha+\beta)(3\alpha^2+3\beta^2-2\alpha\beta(1+\alpha))}{\alpha^2\beta^2}] \bigg{\}} +{\cal A}^2(18\alpha m_s -12m_c) \nonumber\\
&-&18{\cal C}^2 m_s+6{\cal D}^2 m_s +2{\cal AC}\frac{m_c}{\alpha(\alpha-1)}+2{\cal AD}[(3\alpha-1)m_s\nonumber\\
&+&\frac{m_c(4\alpha-3)}{\alpha(\alpha-1)}]-6{\cal BC}[m_s+\frac{m_c}{\alpha(1-\alpha)}]\nonumber\\
&+&2{\cal BD}\frac{m_c(1-4\alpha)+\alpha m_s(-2+5\alpha-3\alpha^2)}{\alpha(\alpha-1)}\bigg{\}}\;,\\
\rho^{\langle \bar{q} q \rangle \langle \bar{s} G s \rangle}(s) &=& \frac{\langle \bar{q} q \rangle \langle \bar{s} G s \rangle}{2^7\pi^2} \int_{\alpha_{min}}^{\alpha_{max}} d \alpha \bigg{\{}\alpha(\alpha-1)(2{\cal AC}+2{\cal BC}+{\cal AD}+{\cal BD}+8{\cal D}^2)\bigg{\}}\;,\\
\rho^{\langle \bar{s} s \rangle \langle \bar{q} G q \rangle}(s) &=& \frac{\langle \bar{s} s \rangle \langle \bar{q} G q \rangle}{2^7\pi^2} \int_{\alpha_{min}}^{\alpha_{max}} d \alpha \bigg{\{}\alpha(\alpha-1)(24{\cal C}^2+8{\cal D}^2)\bigg{\}}\;,\\
\Pi^{\langle \bar{q} q \rangle \langle \bar{s} s \rangle} (M_B^2) &=& \frac{m_c^3\langle \bar{q} q \rangle \langle \bar{s} s \rangle}{3\times 2^6\pi^2} \int_{0}^1 d \alpha \; e^{- \frac{m_c^2}{M_B^2 (1 - \alpha) \alpha}}\bigg{\{} -({\cal AC+AD+BC+BD})(m_q+m_s) \nonumber\\
&+&6({\cal A}^2+{\cal B}^2)[(\alpha-1)m_q-\alpha m_s] \bigg{\}}\;,\\
\Pi^{\langle G^3 \rangle} (M_B^2) &=& \frac{m_c^3\langle g_s^3 G^3 \rangle}{3^2\times 2^{14}\pi^6} \int_{0}^{1} \frac{d \alpha}{\alpha^4} \int_{0}^{1 - \alpha} \frac{d \beta}{\beta^4} \; e^{- \frac{m_c^2 (\alpha + \beta)}{M_B^2 \alpha \beta}}(\alpha+\beta-1)\bigg{\{} 18{\cal A}^2 \beta m_s [\alpha ^5+\alpha ^4 (\beta -1)\nonumber\\
&+&\alpha  \beta ^3 (\beta +2)+\beta ^4 (\beta +1)] +36{\cal B}^2(\alpha^4+\beta^4)+{\cal C}^2(-6 \alpha ^6 m_c\nonumber\\
&-&12 \alpha ^5 \beta  m_c-6 \alpha ^5 m_c-6 \alpha ^4 \beta ^2 m_c-6 \alpha ^4 \beta  m_c+12 \alpha ^4 m_c-6 \alpha ^2 \beta ^4 m_c-12 \alpha  \beta ^5 m_c\nonumber\\
&-&6 \alpha  \beta ^4 m_c-6 \beta ^6 m_c-6 \beta ^5 m_c+12 \beta ^4 m_c)+{\cal D}^2(-18 \alpha ^5 m_c-18 \alpha ^4 \beta  m_c\nonumber\\
&+&18 \alpha ^4 m_c-18 \alpha  \beta ^4 m_c-18 \beta ^5 m_c+18 \beta ^4 m_c)+{\cal AC}[3 m_s (\alpha +\beta ) (\alpha +\beta +1)  (\alpha ^4+\beta ^4 )\nonumber\\
&-&m_c (\alpha +\beta -1)  (\alpha ^4 \beta +(\alpha +2) \alpha ^4+3 \beta ^4 )]+{\cal AD}[3 m_s  (\alpha ^4+\beta ^4 )  (\alpha  (\beta +2)+\beta ^2+\beta  )\nonumber\\
&-&m_c (\alpha +\beta -1)  (3 \alpha ^4+3 \beta ^4 )]+{\cal BC}[6 m_s (\alpha +\beta )  (\alpha ^4+\beta ^4 )\nonumber\\
&-&m_c (\alpha +\beta -1)  (\alpha ^4 \beta +(\alpha +2) \alpha ^4+3 \beta ^4 )]+{\cal BD}[3 m_s  (\alpha ^4+\beta ^4 ) (\alpha  (\alpha +\beta +1)+2 \beta )\nonumber\\
&-&m_c (\alpha +\beta -1)  (3 \alpha ^4+3 \beta ^4 )] \bigg{\}}\;,\\
\Pi^{\langle \bar{q} q\rangle \langle G^2 \rangle} (M_B^2) &=& \frac{m_c^2 \langle \bar{q} q\rangle \langle G^2 \rangle}{3^2\times 2^{11}\pi^4} \int_{0}^{1} d \alpha \bigg{\{} \int_{0}^{1 - \alpha} \frac{d \beta}{\alpha^3\beta^3M_B^2} \; e^{- \frac{m_c^2 (\alpha + \beta)}{M_B^2 \alpha \beta}}\bigg{\{} 12{\cal A}^2[-\alpha  M_B^2  (m_c  (\alpha ^3+\beta ^3 )\nonumber\\
&-&3 \beta  m_s  (\alpha ^2+\beta ^2 ) )-m_c^2 m_s  (\alpha ^3+\beta ^3 )] +12{\cal B}^2[-\alpha  M_B^2  (m_c  (\alpha ^4+\alpha ^3 \beta +\alpha  \beta ^3+\beta ^4 )\nonumber\\
&-&3 \beta  m_s  (\alpha ^2+\beta ^2 ) )-m_c^2 m_s  (\alpha ^3+\beta ^3 )]+12{\cal C}^2[3 \alpha  \beta  m_s (\alpha +\beta )^2 M_B^2\nonumber\\
&-&m_c^2 m_s  (\alpha ^3+\beta ^3 )]+12{\cal D}^2m_s[\alpha  \beta   ((\alpha +3) \alpha ^2+\beta ^2 (\beta +3) ) M_B^2\nonumber\\
&-&m_c^2  (\alpha ^3+\beta ^3 )]+2{\cal AC}[-(\alpha +\beta )^2 M_B^2  (m_c  (\alpha ^2-\alpha  \beta +\beta ^2 )-3 \alpha  \beta  m_s )\nonumber\\
&-&m_c^2 m_s  (\alpha ^3+\beta ^3 )]+2{\cal AD}[M_B^2  (- (m_c (\alpha +\beta )  (\alpha ^2-\alpha  \beta +\beta ^2 )  (\alpha  \beta +\alpha +\beta ^2 )\nonumber\\
&-&\alpha  \beta  m_s  ((\alpha +3) \alpha ^2+\beta ^2 (\beta +3) ) ) ) -m_c^2 m_s  (\alpha ^3+\beta ^3 )] \nonumber\\
&+& 2{\cal BC}[-(\alpha +\beta )^2 M_B^2  (m_c  (\alpha ^3+\beta ^3 )-3 \alpha  \beta  m_s )-m_c^2 m_s  (\alpha ^3+\beta ^3 )]\nonumber\\
&+&2{\cal BD}[M_B^2  (- (m_c (\alpha +\beta )  (\alpha ^2-\alpha  \beta +\beta ^2 ) (\alpha  (\alpha +\beta )+\beta )-\alpha  \beta  m_s  ((\alpha +3) \alpha ^2\nonumber\\
&+&\beta ^2 (\beta +3) ) ) )-m_c^2 m_s  (\alpha ^3+\beta ^3 )]\bigg{\}}+ \frac{e^{- \frac{m_c^2}{M_B^2 (1 - \alpha) \alpha}}}{\alpha^2(\alpha-1)^2} \bigg{\{} ({\cal C}^2+{\cal D}^2) m_s (-12\nonumber\\
&+&36\alpha-36\alpha^2)+2({\cal AC+AD+BC+BD})[1+3\alpha(\alpha-1)] \bigg{\}}\bigg{\}}\;,\\
\Pi^{\langle \bar{s} s\rangle \langle G^2 \rangle} (M_B^2) &=& \frac{m_c^2 \langle \bar{s} s\rangle \langle G^2 \rangle}{3^2\times 2^{11}\pi^4} \int_{0}^{1} d \alpha \bigg{\{} \int_{0}^{1 - \alpha} \frac{d \beta}{\alpha^3\beta^3M_B^2} \; e^{- \frac{m_c^2 (\alpha + \beta)}{M_B^2 \alpha \beta}}\bigg{\{} 6\beta{\cal A}^2M_B^2\left(\alpha ^3+\beta ^3\right)\nonumber\\
&\times& (2 m_c (\alpha +\beta )+\alpha  \text{ms}) +6\beta{\cal B}^2M_B^2\left(\alpha ^3+\beta ^3\right) (2 m_c+\alpha  m_s) \nonumber\\
&+& 6{\cal C}^2m_s \left(m_c^2 \left(\alpha ^4+\alpha ^3 \beta +\alpha  \beta ^3+\beta ^4\right)-\alpha  \beta  \left(4 \alpha ^3+3 \alpha ^2 \beta +3 \alpha  \beta  (\beta +2)+4 \beta ^3\right) M_B^2\right)\nonumber\\
&+&6{\cal D}^2m_s  (m_c^2  (\alpha ^3+\beta ^3 )-\alpha  \beta   ((\alpha +3) \alpha ^2+\beta ^2 (\beta +3) ) M_B^2 )\nonumber\\
&+&{\cal AC}(\alpha +\beta )  (m_c^2 m_s  (\alpha ^3+\beta ^3 )-2 M_B^2  (m_c  (\alpha ^4+\alpha ^3 \beta +\alpha  \beta ^3+\beta ^4 )\nonumber\\
&+&2 \alpha  \beta  m_s  (\alpha ^2+\alpha  \beta +\beta ^2 ) ) ) +{\cal AD}[m_c^2 m_s  (\alpha ^3+\beta ^3 )\nonumber\\
&-&M_B^2  (2 m_c (\alpha +\beta )  (\alpha  \beta +\alpha +\beta ^2 )  (\alpha ^2-\alpha  \beta +\beta ^2 )\nonumber\\
&+&\alpha  \beta  m_s  (\alpha ^3+\alpha ^2 (2 \beta +3)+\beta ^2 (\beta +3) ) )]+{\cal BC}(\alpha +\beta )  (m_c^2 m_s  (\alpha ^3+\beta ^3 )\nonumber\\
&-&2 M_B^2  (m_c  (\alpha ^3+\beta ^3 )+\alpha  \beta  m_s  (2 \alpha ^2+\alpha  \beta +2 \beta ^2 ) ) )+{\cal BD}[m_c^2 m_s  (\alpha ^3+\beta ^3 )\nonumber\\
&-&M_B^2  (2 m_c (\alpha +\beta )  (\alpha ^2-\alpha  \beta +\beta ^2 ) (\alpha  (\alpha +\beta )+\beta )\nonumber\\
&+&\alpha  \beta  m_s  (\alpha ^2 (\alpha +3)+(2 \alpha +3) \beta ^2+\beta ^3 ) )]\bigg{\}}\nonumber\\
& +& \frac{m_s e^{- \frac{m_c^2}{M_B^2 (1 - \alpha) \alpha}}}{\alpha^2(\alpha-1)^2}    \bigg{\{}  {\cal A}^2(-6 \alpha ^3+24 \alpha ^2-18 \alpha +6)  +{\cal B}^2(-6 \alpha ^3+24 \alpha ^2-18 \alpha +6)\nonumber\\
&+&({\cal C}^2+{\cal D}^2)(18 \alpha ^2-18 \alpha +6) +{\cal AC}((\alpha -1) \alpha +1)+{\cal AD}((\alpha -1) \alpha  (8 \alpha -9)-1)\nonumber\\
&+&{\cal BC}(9 (\alpha -1) \alpha +1)+{\cal BD}((\alpha -1) \alpha  (8 \alpha +1)+1)\bigg{\}} \bigg{\}}\;,\\
\Pi^{\langle \bar{q} q \rangle \langle \bar{s} G s \rangle} (M_B^2) &=& \frac{m_c\langle \bar{q} q \rangle \langle \bar{s} G s \rangle}{3^2\times 2^8\pi^2} \int_{0}^1 \frac{d \alpha}{\alpha^2(\alpha-1)^2M_B^4} \; e^{- \frac{m_c^2}{M_B^2 (1 - \alpha) \alpha}}\bigg{\{}18{\cal A}^2 m_c[2 (\alpha -2) (\alpha -1) \alpha  M_B^4\nonumber\\
&+&\alpha  m_c M_B^2 (-2 m_c-\alpha  m_q+m_q)+m_c^3 m_q]+18{\cal B}^2[2 \alpha  M_B^2 (m_c\nonumber\\
&+&(\alpha -1) m_q)  ((\alpha -1) \alpha  M_B^2-m_c^2 )+m_c^4 m_q]+36{\cal C}^2m_c[19 (\alpha -1) \alpha  M_B^2+5 m_c^2]\nonumber\\
&+&36{\cal D}^2m_c[18 (\alpha -1) \alpha  M_B^2+5 m_c^2]-{\cal AC}[(\alpha -1) \alpha  M_B^2  (m_c^2 (12 m_c\nonumber\\
&-&3 m_q-4 m_s)-2 (\alpha -1) \alpha  M_B^2 (15 m_c-2 m_s) )+m_c^4 (3 m_q+2 m_s)]\nonumber\\
&-&{\cal AD}[(\alpha -1) \alpha  M_B^2  (2 (\alpha -1) \alpha  M_B^2 (-12 m_c+3 \alpha  m_q+2 \alpha  m_s)\nonumber\\
&+&m_c^2 (12 m_c-(\alpha +1) (3 m_q+2 m_s)) )+m_c^4 (3 m_q+2 m_s)]\nonumber\\
&+&{\cal BC}[2 (\alpha -1) \alpha  M_B^2  (3 (\alpha -1) \alpha  M_B^2 (5 m_c-m_q)+m_c^2 (-6 m_c+3 m_q+m_s) )\nonumber\\
&-&m_c^4 (3 m_q+2 m_s)]+{\cal BD}[(\alpha -1) \alpha  M_B^2  (2 (\alpha -1) \alpha  M_B^2 (12 m_c+(\alpha -1) (3 m_q+2 m_s))\nonumber\\
&+&m_c^2 (-12 m_c-(\alpha -2) (3 m_q+2 m_s)) )-m_c^4 (3 m_q+2 m_s)]\bigg{\}}\;,\\
\Pi^{\langle \bar{s} s \rangle \langle \bar{q} G q \rangle} (M_B^2) &=& \frac{m_c\langle \bar{s} s \rangle \langle \bar{q} G q \rangle}{ 2^7\pi^2} \int_{0}^1 \frac{d \alpha}{\alpha^2(\alpha-1)^2M_B^4} \; e^{- \frac{m_c^2}{M_B^2 (1 - \alpha) \alpha}}\bigg{\{}-{\cal A}^2[(\alpha -1) \alpha  M_B^2  \nonumber\\
&\times&(2 (\alpha -1) \alpha  M_B^2 (\alpha  m_s-m_c)+m_c^2 (2 m_c+(\alpha -1) m_q-2 \alpha  m_s) )\nonumber\\
&+&m_c^4 (-\alpha  m_q+m_q+\alpha  m_s)]+{\cal B}^2[\alpha  m_c^2 M_B^2  (-2 (\alpha -1) m_c-2 (\alpha -1)^2 m_q\nonumber\\
&+&\alpha  (\alpha +1) m_s )+(\alpha -1) \alpha ^2 M_B^4 (2 (\alpha +1) m_c+(\alpha -1) (2 (\alpha -1) m_q-\alpha  m_s))\nonumber\\
&+&m_c^4 ((\alpha -1) m_q-\alpha  m_s)]+2{\cal C}^2 m_c[5m_c^2+19\alpha(\alpha-1)M_B^2]\nonumber\\
&+&2{\cal D}^2 m_c[5m_c^2+18\alpha(\alpha-1)M_B^2]\bigg{\}}\;,
\end{eqnarray}
where $M_B$ is the Borel parameter introduced by the Borel
transformation, $q = u$ or $d$. Here, we also have the following definitions:
\begin{eqnarray}
{\cal F}_{\alpha \beta} &=& (\alpha + \beta) m_c^2 - \alpha \beta s \; , {\cal H}_\alpha  = m_c^2 - \alpha (1 - \alpha) s \; , \\
\alpha_{min} &=& \left(1 - \sqrt{1 - 4 m_c^2/s} \right) / 2, \; , \alpha_{max} = \left(1 + \sqrt{1 - 4 m_c^2 / s} \right) / 2  \; , \\
\beta_{min} &=& \alpha m_c^2 /(s \alpha - m_c^2).
\end{eqnarray}

\subsection{The spectral densities for tetraquark state}
\begin{eqnarray}
\rho^{pert}(s) &=& \frac{1}{3 \times 2^{10} \pi^6}\int^{\alpha_{max}}_{\alpha_{min}} \frac{d \alpha}{\alpha^3} \int^{1 - \alpha}_{\beta_{min}} \frac{d \beta}{\beta^3} {\cal F}^2_{\alpha \beta} (\alpha + \beta - 1) \bigg{\{} 3 {\cal A}^2 \bigg[{\cal F}_{\alpha \beta}^2(1+\alpha+\beta)- 4m_c {\cal F}_{\alpha \beta} [2 \beta m_s\nonumber\\
  &+&\alpha m_q (1+\alpha+\beta)] + 24 \alpha \beta m_q m_s m_c^2\bigg] + 3  {\cal B}^2 \bigg[{\cal F}_{\alpha \beta}^2(1+\alpha+\beta) -4m_c {\cal F}_{\alpha \beta} [2 \alpha m_s \nonumber\\
  &+&\beta m_q (1+\alpha+\beta)] +24 \alpha \beta m_q m_s m_c^2  \bigg] + 3 \bigg[ ({\cal C}^2+{\cal D}^2)({\cal F}_{\alpha \beta}^2(1+\alpha+\beta)+24 \alpha \beta m_q m_s m_c^2)\nonumber\\
  &+&4  m_c {\cal F}_{\alpha \beta}[\alpha m_q (2{\cal D}^2+{\cal C}^2(1+\alpha+\beta))+\beta m_s (2{\cal C}^2+{\cal D}^2(1+\alpha+\beta))] \bigg]\nonumber\\
  &+&4({\cal AD+BC}){\cal F}_{\alpha \beta}\bigg[6\alpha\beta m_q m_s - (\alpha+\beta-1)^2m_c^2-3 m_c(\alpha+\beta-1)(\alpha m_s-\beta m_q) \bigg]\bigg{\}}\; ,\\
  \rho^{\langle \bar{q} q \rangle}(s) &=& \frac{\langle \bar{q} q \rangle}{2^6\pi^4} \int^{\alpha_{max}}_{\alpha_{min}} d \alpha \bigg{\{} \int^{1 - \alpha}_{\beta_{min}} d \beta \frac{{\cal F}_{\alpha \beta}}{\alpha^2\beta^2} \bigg{\{} {\cal A}^2\bigg[  -2 \alpha^2 m_c {\cal F}_{\alpha \beta} -2 \alpha\beta m_c {\cal F}_{\alpha \beta} -\alpha \beta m_q {\cal F}_{\alpha \beta}\nonumber\\
  &+&4\alpha\beta m_s m_c^2  \bigg]  +{\cal B}^2\bigg[  -2 \alpha m_c {\cal F}_{\alpha \beta} +2 \alpha\beta^2 m_c m_s m_q -\alpha \beta m_q {\cal F}_{\alpha \beta} +4\alpha\beta m_s m_c^2  \bigg]\nonumber\\
  &+&{\cal C}^2\bigg[  2 \alpha^2 m_c {\cal F}_{\alpha \beta} +2 \alpha\beta^2 m_c {\cal F}_{\alpha \beta}  -\alpha \beta m_q {\cal F}_{\alpha \beta} +4\alpha\beta m_s m_c^2  \bigg]\nonumber\\
  &+& {\cal D}^2\bigg[  2 \alpha m_c {\cal F}_{\alpha \beta} -2 \alpha\beta^2 m_c m_s m_q -\alpha \beta m_q {\cal F}_{\alpha \beta} +4\alpha\beta m_s m_c^2  \bigg]\nonumber\\
  &+&{\cal AD}\bigg[ (2\alpha\beta m_q m_c+2\beta {\cal F}_{\alpha \beta})(\alpha m_s +(\alpha+\beta-1)m_c)  \bigg]\nonumber\\
  &+&{\cal BC}\bigg[ (2\alpha\beta m_q m_c-2\beta {\cal F}_{\alpha \beta})(-\alpha m_s +(\alpha+\beta-1)m_c)  \bigg]\bigg{\}}\;, \\
  &-&\frac{m_q {\cal H}_\alpha}{\alpha(\alpha-1)}\bigg[({\cal A}^2+{\cal B}^2+{\cal C}^2+{\cal D}^2){\cal H}_\alpha+2(\alpha-1)m_s m_c({\cal A}^2+{\cal B}^2-{\cal C}^2-{\cal D}^2)\bigg]\bigg{\}}\;,\\
  \rho^{\langle \bar{s} s \rangle}(s) &=& \frac{\langle \bar{s} s \rangle}{2^6\pi^4} \int^{\alpha_{max}}_{\alpha_{min}} d \alpha \bigg{\{} \int^{1 - \alpha}_{\beta_{min}} d \beta \frac{{\cal F}_{\alpha \beta}}{\alpha^2\beta^2} \bigg{\{}- {\cal A}^2\bigg[  2 \beta m_c {\cal F}_{\alpha \beta} -2 \alpha^2\beta m_c m_q m_s +\alpha \beta m_s {\cal F}_{\alpha \beta} \nonumber\\
 & -&4\alpha\beta m_s m_c^2  \bigg]-{\cal B}^2\bigg[  2 \alpha\beta m_c {\cal F}_{\alpha \beta} +2 \beta^2 {\cal F}_{\alpha \beta} +\alpha \beta m_s {\cal F}_{\alpha \beta} -4\alpha\beta m_q m_c^2  \bigg]\nonumber\\
  &-&{\cal C}^2\bigg[  -2 \beta m_c {\cal F}_{\alpha \beta} +2 \alpha^2\beta m_c m_q m_s  +\alpha \beta m_s {\cal F}_{\alpha \beta} -4\alpha\beta m_q m_c^2  \bigg]\nonumber\\
  &-& {\cal D}^2\bigg[ - 2 \alpha\beta m_c {\cal F}_{\alpha \beta} -2 \beta^2 {\cal F}_{\alpha \beta}+\alpha \beta m_s {\cal F}_{\alpha \beta} -4\alpha\beta m_q m_c^2  \bigg]\nonumber\\
  &+&{\cal AD}\bigg[ (2\alpha\beta m_s m_c-2\alpha {\cal F}_{\alpha \beta})(-\beta m_q +(\alpha+\beta-1)m_c)  \bigg]\nonumber\\
  &+&{\cal BC}\bigg[ (2\alpha\beta m_s m_c+2\alpha {\cal F}_{\alpha \beta})(\beta m_q +(\alpha+\beta-1)m_c)  \bigg]\bigg{\}}\;, \\
  &-&\frac{m_q {\cal H}_\alpha}{\alpha(\alpha-1)}\bigg[({\cal A}^2+{\cal B}^2+{\cal C}^2+{\cal D}^2){\cal H}_\alpha-2\alpha m_q m_c({\cal A}^2+{\cal B}^2-{\cal C}^2-{\cal D}^2)\bigg]\bigg{\}}\;,\\
\rho^{\langle G^2 \rangle}(s) &=& \frac{\langle g_s^2 G^2\rangle}{3^2\times2^{12}\pi^6} \int^{\alpha_{max}}_{\alpha_{min}} d \alpha \bigg{\{} \int^{1 - \alpha}_{\beta_{min}} \frac{d \beta}{\alpha^3\beta^3} \bigg{\{} 6{\cal A}^2\bigg[\alpha\beta {\cal F}_{\alpha\beta}^2 ((\alpha + \beta) (3\alpha + \beta) - 2\beta) \nonumber\\
&-& 2 {\cal F}_{\alpha\beta} m_c ( m_c (\alpha + \beta - 1) (\alpha + \beta) (\alpha + \beta + 1) (\alpha^2 - \alpha\beta + \beta^2) + \alpha^2 m_q (-6\alpha^2\beta \nonumber\\
&-& 3\alpha (\alpha^2 - 1)- 2 (\alpha + 1)\beta^2 + \beta^3) + 3\beta^2 m_s (\alpha^2 - 2\alpha\beta - 2 (\beta - 1)\beta )) + 2 m_c^2 (\alpha + \beta \nonumber\\
&-& 1) ( m_c (\alpha^3 + \beta^3) (\alpha m_q (\alpha + \beta + 1) + 2\beta m_s) - 6\alpha\beta m_q m_s (\alpha^2 + \beta^2))\bigg]\nonumber\\
&+&6{\cal B}^2\bigg[\alpha\beta {\cal F}_{\alpha\beta}^2 (4\alpha\beta + (\alpha - 2)\alpha + 3\beta^2) - 2 {\cal F} _ {\alpha\beta} m_c (3\alpha^2 m_q (-2\alpha\beta - 2 (\alpha - 1)\alpha \nonumber\\
&+& \beta^2) + m_c (\alpha + \beta - 1) (\alpha + \beta) (\alpha + \beta + 1) (\alpha^2 - \alpha\beta + \beta^2) + \beta^2 m_s (\alpha^3 \nonumber\\
&-& 2\alpha^2 (\beta + 1) - 6\alpha\beta^2 - 3\beta^3 + 3\beta )) + 2 m_c^2 (\alpha + \beta - 1)(2\alpha m_q ( m_c (\alpha^3 + \beta^3) \nonumber\\
&-& 3\beta m_s (\alpha^2 + \beta^2)) + \beta m_c m_s (\alpha +\beta) (\alpha + \beta + 1) (\alpha^2 - \alpha\beta + \beta^2))  \bigg]\nonumber\\
&+&6{\cal C}^2\bigg[\alpha\beta {\cal F} _ {\alpha\beta}^2 ((\alpha + \beta) (3\alpha + \beta) - 2\beta) - 2 {\cal F} _ {\alpha\beta} m_c (\alpha^2 m_q (6\alpha^2\beta + 3\alpha (\alpha^2 - 1) \nonumber\\
&+& 2 (\alpha + 1)\beta^2 - \beta^3) +m_c (\alpha + \beta - 1) (\alpha + \beta) (\alpha + \beta + 1) (\alpha^2 - \alpha\beta + \beta^2) \nonumber\\
&+& 3\beta^2 m_s (-\alpha^2 + 2\alpha\beta + 2 (\beta - 1)\beta )) - 2 m_c^2 (\alpha + \beta - 1) (\alpha m_q ( m_c (\alpha + \beta) (\alpha \nonumber\\
&+& \beta +1) (\alpha^2- \alpha\beta + \beta^2) + 6\beta m_s (\alpha^2 + \beta^2)) + 2\beta m_Q m_s (\alpha^3 + \beta^3))\bigg]\nonumber\\
&+&6{\cal D}^2\bigg[ \alpha\beta {\cal F}_{\alpha\beta}^2 (4\alpha\beta + (\alpha - 2)\alpha + 3\beta^2) - {\cal F}_{\alpha\beta} m_c (3\alpha^2 m_q (2\alpha\beta + 2 (\alpha - 1)\alpha - \beta^2)\nonumber\\
& +&  m_c (\alpha + \beta - 1) (\alpha + \beta) (\alpha + \beta +   1) (\alpha^2 - \alpha\beta + \beta^2) \nonumber\\
&+& \beta^2 m_s (-\alpha^3 + 2\alpha^2 (\beta + 1) +6\alpha\beta^2 + 3\beta (\beta^2 - 1))) -  m_c^2 (\alpha + \beta - 1) (2\alpha m_q ( m_c (\alpha^3 + \beta^3) \nonumber\\
&+& 3\beta m_s (\alpha^2 + \beta^2)) + \beta m_c m_s (\alpha + \beta) (\alpha + \beta + 1) (\alpha^2 - \alpha\beta + \beta^2))  \bigg] \nonumber\\
&+& {\cal AB}\alpha\beta\bigg[ {\cal F} _ {\alpha\beta} (\beta^3 m_c (4 m_c - 3 m_q) + 3\beta^2 (3 m_c - 2 m_q) (2\alpha m_c + m_c - \alpha m_s) \nonumber\\
&-& 3\beta (  m_q - 2 m_c) ((\alpha (3\alpha + 4) - 3) m_c - 2\alpha (\alpha + 1)m_s)+ (\alpha - 1) m_c ((\alpha (4\alpha + 13) - 5) m_q\nonumber\\
& -& 3\alpha (\alpha + 3) m_s))\bigg]+{\cal AD }\bigg[ m_c (\alpha + \beta - 1) ({\cal F} _ {\alpha\beta} ( m_c (\alpha + \beta - 1) (12 (\alpha - 1)\alpha^2 \nonumber\\
&-& (5\alpha + 12)\beta^2 - 5 (\alpha - 1)\alpha\beta + 12\beta^3) +  36 ((\alpha - 1)\alpha^3 m_s - (\beta - 1)\beta^3 m_q)) \nonumber\\
&+& 4 m_c (\alpha^3 + \beta^3) (  m_c^2 (\alpha + \beta - 1)^2 + 3 m_c (\alpha + \beta - 1) (\alpha m_s - \beta m_q) -  6\alpha\beta m_q m_s))\bigg]\nonumber\\
&+&{\cal BC }\bigg[m_c (\alpha + \beta - 1) ({\cal F} _ {\alpha\beta} ( m_c (\alpha + \beta - 1) (12 (\alpha - 1)\alpha^2 \nonumber\\
&-& (5\alpha + 12)\beta^2 - 5 (\alpha - 1)\alpha\beta + 12\beta^3) +  36 (-(\alpha - 1)\alpha^3 m_s + (\beta - 1)\beta^3 m_q)) \nonumber\\
&+& 4 m_c (\alpha^3 + \beta^3) (  m_c^2 (\alpha + \beta - 1)^2 - 3 m_c (\alpha + \beta - 1) (\alpha m_s - \beta m_q) -  6\alpha\beta m_q m_s))\bigg]\nonumber\\
&+& {\cal CD}\alpha\beta\bigg[ {\cal F} _ {\alpha\beta} (\beta^3 m_c (4 m_c + 3 m_q) + 3\beta^2 (3 m_c+ 2 m_q) (2\alpha m_c + m_c + \alpha m_s) \nonumber\\
&+& 3\beta (  m_q + 2 m_c) ((\alpha (3\alpha + 4) - 3) m_c + 2\alpha (\alpha + 1)m_s)+ (\alpha - 1) m_c ((\alpha (4\alpha + 13) - 5) m_c\nonumber\\
& +& 3\alpha (\alpha + 3) m_s))\bigg]\bigg{\}}\bigg{\}}\;,\\
\rho^{\langle \bar{q} G q \rangle}(s) &=& \frac{\langle g_s \bar{q} \sigma \cdot G q \rangle}{3\times2^8\pi^4} \int_{\alpha_{min}}^{\alpha_{max}} \bigg{\{} \int_{\beta_{min}}^{1 - \alpha} d \beta\frac{1}{\alpha^2\beta^2} \bigg{\{} 12{\cal A}^2 \alpha^2 m_c [(\alpha+2\beta){\cal F}_{\alpha\beta}-\beta m_c m_s]\nonumber\\
&-&12{\cal C}^2 \alpha^2 m_c [(\alpha+2\beta){\cal F}_{\alpha\beta}+\beta m_c m_s] + {\cal AB} \beta^2 {\cal F}_{\alpha\beta}[(2\alpha +\beta-1)m_c-\alpha m_s]\nonumber\\
 &-&{\cal CD} \beta^2 {\cal F}_{\alpha\beta}[(2\alpha +\beta+1)m_c+\alpha m_s]+{\cal AD}\beta[6\alpha{\cal F}_{\alpha\beta}(m_c-\alpha m_c-\alpha m_s)\nonumber\\
 &-&m_c({\cal F}_{\alpha\beta}(1+19\alpha)+4\alpha^2 m_q m_c)-\beta^2 m_c {\cal F}_{\alpha\beta}]+{\cal BC}\beta[-6\alpha{\cal F}_{\alpha\beta}(m_c-\alpha m_c+\alpha m_s)\nonumber\\
 &+&m_c({\cal F}_{\alpha\beta}(1+19\alpha)-4\alpha^2 m_q m_c)+\beta^2 m_c {\cal F}_{\alpha\beta}]\bigg{\}}+\frac{4}{\alpha-1}\bigg{\{}{\cal A}^2[{\cal H}_\alpha(3m_c+2m_q(\alpha-1))\nonumber\\
 &+&m_c(\alpha-1)(-m_c m_q+3m_c m_s+m_s m_q(\alpha-1))]+{\cal B}^2[{\cal H}_\alpha(3m_c+2m_q(\alpha-1))\nonumber\\
 &+&m_c(\alpha-1)(-m_c m_q+3m_c m_s+2m_s m_q(\alpha-1))]+{\cal C}^2[{\cal H}_\alpha(-3m_c+2m_q(\alpha-1))\nonumber\\
 &-&m_c(\alpha-1)(m_c m_q-3m_c m_s+2m_s m_q(\alpha-1))]+{\cal D}^2[{\cal H}_\alpha(-3m_c+2m_q(\alpha-1))\nonumber\\
 &-&m_c(\alpha-1)(m_c m_q-3m_c m_s+m_s m_q(\alpha-1))]+{\cal AD}m_s(\alpha-1)(3{\cal H}_\alpha+\alpha m_c m_q)\nonumber\\
& +&{\cal BC}m_s(\alpha-1)(3{\cal H}_\alpha-\alpha m_c m_q)+2{\cal H}_\alpha[{\cal AB}(-m_c+ms)+{\cal CD}(m_c+m_s)]\bigg{\}}\bigg{\}}\;,\\
\rho^{\langle \bar{s} G s \rangle}(s) &=& \frac{\langle g_s \bar{s} \sigma \cdot G s \rangle}{3\times2^8\pi^4} \int_{\alpha_{min}}^{\alpha_{max}} \bigg{\{} \int_{\beta_{min}}^{1 - \alpha} d \beta\frac{1}{\alpha^2\beta^2} \bigg{\{} 12{\cal B}^2 \beta^2 m_c [(2\alpha+\beta){\cal F}_{\alpha\beta}-\alpha m_c m_s]\nonumber\\
&-&12{\cal D}^2 \beta^2 m_c [(2\alpha+\beta){\cal F}_{\alpha\beta}+\alpha m_c m_s] + {\cal AB} \alpha^2 {\cal F}_{\alpha\beta}[(\alpha +2\beta+1)m_c-\beta m_s]\nonumber\\
 &-&{\cal CD} \alpha^2 {\cal F}_{\alpha\beta}[(\alpha +2\beta+1)m_c+\alpha m_s]+{\cal AD}\alpha[-6\beta^2{\cal F}_{\alpha\beta}m_q -4 \alpha\beta^2 m_c^2 m_s\nonumber\\
 &-&m_c{\cal F}_{\alpha\beta}(\alpha+\alpha^2+19\alpha\beta+6\beta^2-6\beta)]-{\cal BC}\alpha[6\beta^2{\cal F}_{\alpha\beta}m_q +4 \alpha\beta^2 m_c^2 m_s\nonumber\\
 &+&m_c{\cal F}_{\alpha\beta}(\alpha+\alpha^2+19\alpha\beta+6\beta^2-6\beta)]\bigg{\}}+\frac{4}{\alpha}\bigg{\{}{\cal A}^2[{\cal H}_\alpha(-3m_c+2m_q\alpha)\nonumber\\
 &+&m_c\alpha(-m_c m_s+3m_c m_q-m_s m_q\alpha)]+{\cal B}^2[{\cal H}_\alpha(-3m_c+2m_q\alpha)\nonumber\\
 &+&m_c\alpha(-m_c m_s+3m_c m_q-2m_s m_q\alpha)]+{\cal C}^2[{\cal H}_\alpha(3m_c+2m_q\alpha)\nonumber\\
 &+&m_c\alpha(-m_c m_s+3m_c m_s+m_s m_q\alpha)]+{\cal D}^2[{\cal H}_\alpha(3m_c+2m_q\alpha)\nonumber\\
 &+&m_c\alpha(-m_c m_s+3m_c m_s+2m_s m_q\alpha)]+{\cal AD}m_q\alpha(3{\cal H}_\alpha+(\alpha-1) m_c m_q)\nonumber\\
& +&{\cal BC}m_q\alpha(3{\cal H}_\alpha-(\alpha-1) m_c m_s)+2{\cal H}_\alpha[{\cal AB}(-m_c+ms)+{\cal CD}(m_c+m_s)]\bigg{\}}\bigg{\}}\;,\\
\rho^{\langle \bar{q} q \rangle \langle \bar{s} s \rangle}(s) &=& \frac{\langle \bar{q} q \rangle \langle \bar{s} s \rangle}{3\times2^4\pi^2} \int_{\alpha_{min}}^{\alpha_{max}} d \alpha \bigg{\{}{\cal A}^2[2m_c(2m_c-\alpha m_s)+(\alpha-1)m_q(4m_c-3\alpha m_s)]\nonumber\\
&+&{\cal B}^2[4m_c(m_c-\alpha m_s)+(\alpha-1)m_q(2m_c-3\alpha m_s)]\nonumber\\
&+&{\cal C}^2[2m_c(2m_c+\alpha m_s)-(\alpha-1)m_q(4m_c+3\alpha m_s)]\nonumber\\
&+&{\cal D}^2[4m_c(m_c+\alpha m_s)-(\alpha-1)m_q(2m_c+3\alpha m_s)]\nonumber\\
&+&{\cal AD}[4{\cal H}_\alpha +2 m_c m_s(\alpha-1)+2m_c m_q \alpha]\nonumber\\
&+&{\cal BC}[4{\cal H}_\alpha - 2m_c m_s(\alpha-1)-2m_c m_q \alpha]\bigg{\}}\;,\\
\rho^{\langle G^3 \rangle}(s) &=& \frac{\langle g_s^3 G^3 \rangle}{3\times2^{12}\pi^6} \int_{\alpha_{min}}^{\alpha_{max}} d\alpha \int_{\beta_{min}}^{1 - \alpha} \frac{d \beta}{\beta^3} (\alpha+\beta-1)\bigg{\{} {\cal A}^2\bigg[2{\cal F}_{\alpha\beta}(\alpha+\beta+1)\nonumber\\
&+&m_c[4\alpha m_c(\alpha+\beta+1)-2m_s(6\alpha+\beta)-m_q(6\alpha^2+6\alpha+7\alpha\beta+\beta+\beta^2)]\bigg]\nonumber\\
&+&{\cal B}^2\bigg[2{\cal F}_{\alpha\beta}(\alpha+\beta+1)+m_c[4\alpha m_c(\alpha+\beta+1)-2m_q(6\alpha+\beta)\nonumber\\
&-&m_s(6\alpha^2+6\alpha+7\alpha\beta+\beta+\beta^2)]\bigg]+{\cal C}^2\bigg[2{\cal F}_{\alpha\beta}(\alpha+\beta+1)\nonumber\\
&+&m_c[4\alpha m_c(\alpha+\beta+1)+2m_s(6\alpha+\beta)+m_q(6\alpha^2+6\alpha+7\alpha\beta+\beta+\beta^2)]\bigg]\nonumber\\
&+&{\cal D}^2\bigg[2{\cal F}_{\alpha\beta}(\alpha+\beta+1)+m_c[4\alpha m_c(\alpha+\beta+1)+2m_q(6\alpha+\beta)\nonumber\\
&+&m_s(6\alpha^2+6\alpha+7\alpha\beta+\beta+\beta^2)]\bigg] +\bigg[ 4({\cal AD}+{\cal BC})[\alpha\beta m_q m_s-m_c^2(\alpha+\beta-1)^2]\nonumber\\
&+&({\cal AD}-{\cal BC})m_c(m_q-m_s)(6\alpha^2-6\alpha+7\alpha\beta+\beta^2-\beta)  \bigg]\bigg{\}}\;,\\
\rho^{\langle \bar{q} q \rangle \langle \bar{s} G s \rangle}(s) &=& \frac{\langle \bar{q} q \rangle \langle \bar{s} G s \rangle}{3\times2^6\pi^2} \int_{\alpha_{min}}^{\alpha_{max}} d \alpha \bigg{\{}\alpha({\cal AB}+{\cal CD})-2({\cal AD}+{\cal BC})(6\alpha^2-7\alpha-1)\bigg{\}}\;,\\
\rho^{\langle \bar{s} s \rangle \langle \bar{q} G q \rangle}(s) &=& \frac{\langle \bar{s} s \rangle \langle \bar{q} G q \rangle}{3\times2^6\pi^2} \int_{\alpha_{min}}^{\alpha_{max}} d \alpha \bigg{\{}(1-\alpha)({\cal AB}+{\cal CD})-2({\cal AD}+{\cal BC})(6\alpha^2-5\alpha)\bigg{\}}\;,\\
\Pi^{\langle G^2 \rangle} (M_B^2) &=& \frac{m_c^4 m_q m_s\langle g_s^2 G^2 \rangle}{3\times 2^{9}\pi^6} \int_{0}^{1} \frac{d \alpha}{\alpha^3} \int_{0}^{1 - \alpha} \frac{d \beta}{\beta^3} \; e^{- \frac{m_c^2 (\alpha + \beta)}{M_B^2 \alpha \beta}}({\cal A}^2+{\cal B}^2+{\cal C}^2+{\cal D}^2)\nonumber\\
&\times&[\alpha^4+\alpha^3(\beta-1)+\alpha\beta^3+\beta^3(\beta-1)]\;,\\
\Pi^{\langle \bar{q} G q \rangle}(M_B^2) &=& \frac{m_c^3 m_q m_s\langle g_s \bar{q} \sigma \cdot G q \rangle}{3\times2^6\pi^4} \int_{0}^{1} \bigg{\{} -\frac{e^{ \frac{m_c^2}{M_B^2 \alpha(\alpha-1)}}}{\alpha}({\cal A}^2+{\cal B}^2-{\cal C}^2-{\cal D}^2) \bigg{\}}\;,\\
\Pi^{\langle \bar{s} G s \rangle}(M_B^2) &=& \frac{m_c^3 m_q m_s\langle g_s \bar{s} \sigma \cdot G s \rangle}{3\times2^6\pi^4} \int_{0}^{1} \bigg{\{} \frac{e^{ \frac{m_c^2}{M_B^2 \alpha(\alpha-1)}}}{\alpha-1}({\cal A}^2+{\cal B}^2-{\cal C}^2-{\cal D}^2) \bigg{\}}\;,\\
\Pi^{\langle \bar{q} q\rangle\langle\bar{s} s \rangle}(M_B^2) &=& \frac{m_c^2 \langle \bar{q} q\rangle\langle\bar{s} s \rangle}{3\times2^4\pi^2} \int_{0}^{1}\frac{e^{ \frac{m_c^2}{M_B^2 \alpha(\alpha-1)}}}{\alpha(\alpha-1)} \bigg{\{} ({\cal A}^2+{\cal B}^2)[2\alpha m_c m_s +m_q (\alpha-1)(-2m_c + 3\alpha m_s)]\nonumber\\
&+&({\cal C}^2+{\cal D}^2)[-2\alpha m_c m_s +m_q (\alpha-1)(2m_c + 3\alpha m_s)]\nonumber\\
&+&{\cal AD+BC}m_q m_s \alpha(\alpha-1) \bigg{\}}\;,\\
\Pi^{\langle G^3 \rangle} (M_B^2) &=& \frac{m_c^2\langle g_s^3 G^3 \rangle}{3^2\times 2^{11}\pi^6} \int_{0}^{1} d \alpha \int_{0}^{1 - \alpha} \frac{d \beta}{M_B^2\beta^4} \; e^{- \frac{m_c^2 (\alpha + \beta)}{M_B^2 \alpha \beta}}(\alpha+\beta-1)\bigg{\{}{\cal A}^2[6(\alpha+\beta)m_c m_s M_B^2\nonumber\\
&+&3m_q(4m_c^2 m_s-12\beta m_s M_B^2+m_c M_B^2(\alpha+\beta+1)(\alpha+\beta))]+{\cal B}^2[6(\alpha+\beta)\nonumber\\
&\times&m_c m_q M_B^2+3m_s(4m_c^2 m_q-12\beta m_q M_B^2+m_c M_B^2(\alpha+\beta+1)(\alpha+\beta))]\nonumber\\
 &-&{\cal C}^2[6(\alpha+\beta)m_c m_s M_B^2+3m_q(-4m_c^2 m_s+12\beta m_s M_B^2+m_c M_B^2(\alpha+\beta\nonumber\\
 &+&1)(\alpha+\beta))] -{\cal D}^2[6(\alpha+\beta)m_c m_q M_B^2+3m_s(-4m_c^2 m_q+12\beta m_q M_B^2\nonumber\\
 &+&m_c M_B^2(\alpha+\beta+1)(\alpha+\beta))] +({\cal AD+BC})M_B^2[-12\alpha\beta m_q m_s +m_c^2(\alpha+\beta-1)^2]\nonumber\\
 &+&3({\cal AD-BC})M_B^2 m_c(m_q-m_s)(\alpha+\beta)(\alpha+\beta-1)\bigg{\}}\;,\\
 \Pi^{\langle \bar{q} q \rangle \langle \bar{s} G s \rangle} (M_B^2) &=& \frac{m_c\langle \bar{q} q \rangle \langle \bar{s} G s \rangle}{3^2\times 2^7\pi^2} \int_{0}^1 \frac{d \alpha}{\alpha^2(\alpha-1)^3M_B^6} \; e^{- \frac{m_c^2}{M_B^2 (1 - \alpha) \alpha}}\bigg{\{}-4{\cal A}^2(\alpha-1)[m_c m_s\nonumber\\
 &+&3(\alpha-1)M_B^2] \bigg[ 2\alpha m_c M_B^2 (-m_c^2+\alpha (\alpha-1)M_B^2) +m_q(m_c^4 -2\alpha(\alpha-1) m_c^2 M_B^2\nonumber\\
 &+&2\alpha^2(\alpha-1)^2 M_B^4) \bigg]+4{\cal B}^2(\alpha-1)(-m_c m_q+2\alpha M_B^2)\bigg[m_c^4 m_s+m_c^2 M_B^2 (\alpha\nonumber\\
 &-&1)(3m_c-2\alpha m_s) +(\alpha-1)^2 M_B^2 (-3 m_c (\alpha-1) +2 \alpha^2 m_s) \bigg] \nonumber\\
&+&4{\cal C}^2(\alpha-1)[-m_c m_s +3(\alpha-1)M_B^2] \bigg[ 2\alpha m_c M_B^2 (m_c^2-\alpha (\alpha-1)M_B^2) \nonumber\\
&+&m_q(m_c^4 -2\alpha(\alpha-1) m_c^2 M_B^2+2\alpha^2(\alpha-1)^2 M_B^4) \bigg] \nonumber\\
&-&4{\cal D}^2(\alpha-1)(m_c m_q+2\alpha M_B^2)\bigg[m_c^4 m_s-m_c^2 M_B^2 (\alpha\nonumber\\
 &-&1)(3m_c+2\alpha m_s) +(\alpha-1)^2 M_B^2 (3 m_c (\alpha-1) +2 \alpha^2 m_s) \bigg] \nonumber\\
 &+&{\cal AB}\alpha^2(\alpha-1)M_B^4[-2m_q m_c^2+M_B^2(\alpha-1)(-4m_c+\alpha m_q)]\nonumber\\
 &+&{\cal CD}\alpha^2(\alpha-1)M_B^4[2m_q m_c^2-M_B^2(\alpha-1)(4m_c+\alpha m_q)]\nonumber\\
 &+&{\cal AD} \alpha (\alpha-1)M_B^2\bigg[ 4(\alpha-1)m_c^3 m_q m_s +2M_B^2 m_c( 4(\alpha-1)^2 m_c m_s \nonumber\\
 &-&2\alpha(\alpha-1)^2m_q m_s +\alpha(6\alpha-7)m_q m_c) -\alpha(\alpha-1)M_B^4((\alpha (24\alpha - 31)\nonumber\\
 &+& 6) m_q + 8 (\alpha - 1) (2 (\alpha - 1) m_s - 3 m_c)) \bigg]\nonumber\\
 &+&{\cal BC} \alpha (\alpha-1)M_B^2\bigg[ 4(\alpha-1)m_c^3 m_q m_s -2M_B^2 m_c( 4(\alpha-1)^2 m_c m_s \nonumber\\
 &+&2\alpha(\alpha-1)^2m_q m_s +\alpha(6\alpha-7)m_q m_c) +\alpha(\alpha-1)M_B^4((\alpha (24\alpha - 31)\nonumber\\
 &+& 6) m_q + 8 (\alpha - 1) (2 (\alpha - 1) m_s + 3 m_c)) \bigg]   \bigg{\}}\;,\\
 \Pi^{\langle \bar{s} s \rangle \langle \bar{q} G q \rangle} (M_B^2) &=& \frac{m_c\langle \bar{s} s \rangle \langle \bar{q} G q \rangle}{3^2\times 2^7\pi^2} \int_{0}^1 \frac{d \alpha}{\alpha^2(\alpha-1)^3M_B^6} \; e^{- \frac{m_c^2}{M_B^2 (1 - \alpha) \alpha}}\bigg{\{}-4{\cal A}^2(\alpha-1)[m_c m_s\nonumber\\
 &+&2(\alpha-1)M_B^2] \bigg[ 3\alpha m_c M_B^2 (-m_c^2+\alpha^2 M_B^2) +m_q(m_c^4 -2\alpha(\alpha-1) m_c^2 M_B^2\nonumber\\
 &+&2\alpha^2(\alpha-1)^2 M_B^4) \bigg]+4{\cal B}^2(\alpha-1)(-m_c m_q+3\alpha M_B^2)\bigg[m_c^4 m_s+m_c^2 M_B^2 (\alpha\nonumber\\
 &-&1)(m_c-\alpha m_s) +2\alpha(\alpha-1)^2 M_B^2 (- m_c + \alpha m_s) \bigg] \nonumber\\
&+&4{\cal C}^2(\alpha-1)[-m_c m_s +2(\alpha-1)M_B^2] \bigg[ 3\alpha m_c M_B^2 (m_c^2-\alpha^2 M_B^2) \nonumber\\
&+&m_q(m_c^4 -2\alpha(\alpha-1) m_c^2 M_B^2+2\alpha^2(\alpha-1)^2 M_B^4) \bigg] \nonumber\\
&-&4{\cal D}^2(\alpha-1)(m_c m_q+3\alpha M_B^2)\bigg[m_c^4 m_s-2m_c^2 M_B^2 (\alpha\nonumber\\
 &-&1)3m_c+\alpha m_s) +2\alpha(\alpha-1)^2 M_B^2 ( m_c \alpha + \alpha m_s) \bigg] \nonumber\\
 &+&{\cal AB}(\alpha-1)^3 M_B^4[-2m_q m_c^2+M_B^2\alpha(4m_c+(\alpha-1) m_s)]\nonumber\\
 &+&{\cal CD}(\alpha-1)^3 M_B^4[2m_q m_c^2+M_B^2\alpha(4m_c+-(\alpha-1) m_s)]\nonumber\\
 &+&{\cal AD}  (\alpha-1)^2M_B^2\bigg[ 4\alpha m_c^3 m_q m_s +2M_B^2 m_c( 4\alpha^2 m_c m_s \nonumber\\
 &-&2\alpha^2(\alpha-1) m_q m_s +(\alpha-1)(6\alpha+1)m_s m_c) -\alpha(\alpha-1)M_B^4((\alpha (24\alpha - 17)\nonumber\\
 &-& 1) m_s + 8 \alpha  (2 \alpha  m_s - 3 m_c)) \bigg]\nonumber\\
 &+&{\cal BC}  (\alpha-1)^2M_B^2\bigg[ 4\alpha m_c^3 m_q m_s -2M_B^2 m_c( 4\alpha^2 m_c m_s \nonumber\\
 &+&2\alpha^2(\alpha-1) m_q m_s +(\alpha-1)(6\alpha+1)m_s m_c) +\alpha(\alpha-1)M_B^4((\alpha (24\alpha - 17)\nonumber\\
 &-& 1) m_s + 8 \alpha  (2 \alpha  m_s + 3 m_c)) \bigg]  \bigg{\}}\;,
\end{eqnarray}

\section{The decay spectral densities of $Z^+_{cs}$}
\subsection{The decay spectral densities of $Z^+_{cs}$ for molecular state}

To calculate the decay spectral densities of $Z^+_{cs}$ in molecular structure, we isolate the $g_{\mu\nu}$ structure of both side of 
Eqs.(\ref{three-point-p}) and (\ref{three-point-o}).
On the OPE side of QCD sum rules for the current (\ref{current-mixing1}), the three-point function of $Z^+_{cs} \to \bar{D}^\ast D_s^+$ after Borel transformation may write: 
\begin{eqnarray}
\Pi^{pert}(s_0,v_0,M_{1B}^2,M_{2B}^2) &=& \frac{9 {\cal A}}{8\pi^2} \int_{m_c^2}^{s_0} d s \int_{m_c^2}^{v_0} d v \int_0^\Lambda d \alpha \int_0^\Lambda d \beta\nonumber\\
&\times& e^{-\frac{s}{M_{1B}^2}-\frac{v}{M_{2B}^2}}[m_c^2-(1-\alpha) s][m_c^2-(1-\beta)v]\;,\\
\Pi^{\langle\bar{q}q\rangle}(s_0,v_0,M_{1B}^2,M_{2B}^2) &=&\frac{3 m_c \langle\bar{q}q\rangle {\cal A} e^{\frac{-m_c^2}{M_{2B}^2}} }{\pi^2} \int_{m_c^2}^{s_0} d s \int_0^\Lambda d \alpha  \; e^{-\frac{s}{M_{1B}^2}}[m_c^2-(1-\alpha) s]\;,\\
\Pi^{\langle\bar{s}s\rangle}(s_0,v_0,M_{1B}^2,M_{2B}^2) &=&\frac{3 m_c \langle\bar{s}s\rangle {\cal A} e^{\frac{-m_c^2}{M_{1B}^2}} }{2 \pi^2} \int_{m_c^2}^{v_0} d v \int_0^\Lambda d \beta \; e^{-\frac{v}{M_{2B}^2}} [m_c^2-(1-\beta) v]\;,\\
\Pi^{\langle G^2 \rangle}(s_0,v_0,M_{1B}^2,M_{2B}^2) &=&-\frac{m_c^2 \langle G^2 \rangle {\cal A}}{64\pi^4}\bigg{\{} \int_{m_c^2}^{s_0} d s  \int_0^\Lambda d \alpha \int_0^1 d \beta\ \;e^{-\frac{s}{M_{1B}^2}-\frac{m_c^2}{(1-\beta)M_{2B}^2}}\nonumber\\
&\times&\frac{(m_c^2-(1-\alpha)s)}{{(1-\beta)^2M_{2B}^2}}+\int_{m_c^2}^{v_0} d v  \int_0^1 d \alpha \int_0^\Lambda d \beta\ \;e^{-\frac{v}{M_{2B}^2}-\frac{m_c^2}{(1-\alpha)M_{1B}^2}}\nonumber\\
&\times&\frac{(m_c^2-(1-\beta)v)}{{(1-\alpha)^2M_{1B}^2}}\bigg{\}}\;,\\
\Pi^{\langle \bar{q} G q \rangle}(s_0,v_0,M_{1B}^2,M_{2B}^2)&=&\frac{-3m_c^3 \langle \bar{q} G q \rangle {\cal A}\;e^{\frac{-m_c^2}{M_{2B}^2}}}{4\pi^2 M_{2B}^4}\int_{m_c^2}^{s_0} d s \int_0^\Lambda d \alpha  e^{-\frac{s}{M_{1B}^2}} [m_c^2-(1-\alpha) s]\;,\\
\Pi^{\langle \bar{s} G s \rangle}(s_0,v_0,M_{1B}^2,M_{2B}^2)&=&\frac{-3m_c^3 \langle \bar{q} G q \rangle {\cal A}\;e^{\frac{-m_c^2}{M_{1B}^2}}}{8\pi^2 M_{2B}^4}\int_{m_c^2}^{v_0} d v \int_0^\Lambda d \beta \; e^{-\frac{v}{M_{1B}^2}} [m_c^2-(1-\beta) v]\;,\\
\Pi^{\langle\bar{q}q\rangle\langle\bar{s}s\rangle}(s_0,v_0,M_{1B}^2,M_{2B}^2) &=&4 m_c^2 {\cal A} \langle\bar{q}q\rangle\langle\bar{s}s\rangle e^{-\frac{m_c^2}{M_{1B}^2}-\frac{m_c^2}{M_{1B}^2}}\;,
\end{eqnarray}
where $s_0$ and $v_0$ are the continuum threshold of $D_s$ and $\bar{D}^\ast$, and $M_{1B}^2$ and $M_{2B}^2$ are their Borel parameters, respectively, and $\Lambda=(1-m_c^2/s)$. Since $m_{D_s}\approx m_{\bar{D}^\ast}$, we can set $s_0=v_0$ and $M_{1B}^2=M_{2B}^2\equiv M_B^2$.
 After Borel transformation on the Eq.(\ref{three-point-p}), the phenomenological side of three-point is then obtained as
 \begin{eqnarray}\label{three-point-p-B}
   \Pi^{phen}(s_0,M_B^2)&=& \frac{3\lambda_{Z^+_{cs}} m_{D^\ast} f_{D^\ast} f_{D_s} m_{D_s}^2 g_{Z^+_{cs} \bar{D}^\ast D_s} }{4(m_c+m_s)(m_{Z^+_{cs}}^2/4-m_{D^\ast}^2)} \;  \nonumber\\
      &\times&(e^{-m^2_{D^\ast}/M_B^2}-e^{-m^2_{Z^+_{cs}}/(4 M_B^2)})e^{-m^2_{D_s}/M_B^2}\; .
\end{eqnarray}

In our calculation, $\sqrt{s_0} = 2.1\; \rm{GeV}$ and $1.5\; \rm{GeV}^2 \le M_B^2 \le 2.5\; \rm{GeV}^2$ is the proper Borel window for decay process.

The OPE side of the three-point function of  the decay process $Z^+_{cs} \to \bar{D} D_s^\ast$ is just the transformation of three-point function of  the decay process $Z^+_{cs} \to \bar{D}^\ast D_s^+$ with $\langle\bar{q}q\rangle \leftrightarrow \langle\bar{s}s\rangle$, $\langle \bar{q} G q \rangle \leftrightarrow \langle \bar{s} G s \rangle$, and $\cal{A} \to \cal{B}$, and the phenomenological side of the three-point function of  $Z^+_{cs} \to \bar{D} D_s^\ast$ is just the transformation of Eq.(\ref{three-point-p-B}) with $D_s^+ \to \bar{D}$ and $\bar{D}^\ast \to D_s^\ast$. In numerical analysis, $\sqrt{s_0} = 2.5\; \rm{GeV}$ and $1.7\; \rm{GeV}^2 \le M_B^2 \le 2.3\; \rm{GeV}^2$ is the proper Borel window for decay process.

The OPE side of the three-point function of the decay process $Z^+_{cs} \to J/\psi K^+$ reads as
\begin{eqnarray}
\Pi^{pert}(s_0,v_0,M_{1B}^2,M_{2B}^2) &=&\frac{9 \cal{C}}{16 \pi^4} \int_{4m_c^2}^{s_0} d s \int_{m_s^2}^{v_0} d v \int_{\alpha_{min}}^{\alpha_{max}} d \alpha\; e^{-\frac{s}{M_{1B}^2}-\frac{v}{M_{2B}^2}} (m_c^2-{\cal H}_\alpha) v \;,\\
\Pi^{\langle\bar{q}q\rangle}(s_0,v_0,M_{1B}^2,M_{2B}^2) &=&\frac{9 m_s \langle\bar{q}q\rangle {\cal C}}{2 \pi^2} \int_{4m_c^2}^{s_0} d s \int_{\alpha_{min}}^{\alpha_{max}} d \alpha\; e^{-\frac{s}{M_{1B}^2}}({\cal H}_\alpha - m_c^2)\;,\\
\Pi^{\langle\bar{s}s\rangle}(s_0,v_0,M_{1B}^2,M_{2B}^2) &=&\frac{9 m_s \langle\bar{s}s\rangle {\cal C}}{4 \pi^2} \int_{4m_c^2}^{s_0} d s \int_{\alpha_{min}}^{\alpha_{max}} d \alpha\; e^{-\frac{s}{M_{1B}^2}}( m_c^2-{\cal H}_\alpha)\;,\\
\Pi^{\langle G^2 \rangle}(s_0,v_0,M_{1B}^2,M_{2B}^2) &=&\frac{\langle G^2 \rangle {\cal C}}{128 \pi^4} \bigg{\{}\int_{m_s^2}^{v_0} d v \int_0^1 d \alpha \; e^{-\frac{m_c^2}{\alpha(1-\alpha)M_{1B}^2}-\frac{v}{M_{2B}^2}} \bigg( \frac{3 v (m_c^2-\alpha(\alpha-1)M_{1B}^2)}{128\alpha(\alpha-1)M_{1B}^2}\nonumber\\
&+&\frac{3v m_c^2[m_c^2(1-3\alpha+3\alpha^2)+2\alpha M_{1B}^2 (2\alpha^3-4\alpha^2+3\alpha-1)]}{128\alpha^3(\alpha-1)^3M_{1B}^4} \bigg)\nonumber\\
&+& \int_{4m_c^2}^{s_0} d s \int_{\alpha_{min}}^{\alpha_{max}} d \alpha 18 e^{\frac{s}{M_{1B}^2}}(m_c^2-{\cal H}_\alpha) \bigg{\}}\;,\\
\Pi^{\langle G^3 \rangle}(s_0,v_0,M_{1B}^2,M_{2B}^2) &=&\frac{\langle G^3 \rangle{\cal C}}{512\pi^4}\int_{m_s^2}^{v_0} d v \int_0^1 d \alpha \frac{v m_c^2\; e^{-\frac{m_c^2}{\alpha(1-\alpha)M_{1B}^2}-\frac{v}{M_{2B}^2}}}{\alpha^4(\alpha-1)^4M_{1B}^6} [6m_c^2(2 \alpha ^4\nonumber\\
&-&4 \alpha ^3+6 \alpha ^2-4 \alpha +1) + \alpha M_{1B}^2 (4 \alpha ^5-12 \alpha ^4\nonumber\\
&+&59 \alpha ^3-98 \alpha ^2+62 \alpha -15) ]\;,
\end{eqnarray}
where $s_0$ and $v_0$ are the continuum threshold of $J/\psi$ and $K^+$, and $M_{1B}^2$ and $M_{2B}^2$ are their Borel parameters, respectively. After employing Borel transformation to Eq.(\ref{three-point-p}) with $\bar{D}^\ast \to J/\psi$ and $D_s \to K^+$, the phenomenological side of three-point function of $Z^+_{cs} \to J/\psi K^+$ will be obtained.
\begin{eqnarray}
\Pi^{phen}(s_0,v_0,M_{1B}^2,M_{2B}^2)&=&\frac{3 \lambda_{Z_{cs}^+}m_{J/\psi}f_{J/\psi}f_{K^+}m_{K^+}^2g_{Z_{cs}^+J/\psi K^+}}{m_s(m_{Z_{cs}^+}^2-m_{J/\psi}^2)}\nonumber\\
&\times&(e^{-m_{Z_{cs}}^2/M_{1B}^2}-e^{-m_{J/\psi}^2/M_{1B}^2})e^{-m_{K^+}^2/M_{2B}^2}\;.\label{three-point-p-BJ}
\end{eqnarray}
In numerical analysis, $\sqrt{s_0} = 3.2\; \rm{GeV}$, $\sqrt{v_0} = 0.6\; \rm{GeV}$, $2.0\; \rm{GeV}^2 \le M_{1B}^2 \le 3.0\; \rm{GeV}^2$, and $2.0\; \rm{GeV}^2 \le M_{2B}^2 \le 3.0\; \rm{GeV}^2$ is the proper Borel window for decay process. Here we set $v_0=(m_K+m_{\pi})^2$.

The OPE side of the three-point function of  the decay process $Z^+_{cs} \to\eta_c K^{+\ast}$ reads:
\begin{eqnarray}
\Pi^{pert}(s_0,v_0,M_{1B}^2,M_{2B}^2) &=&\frac{3 \cal{D}}{16 \pi^4} \int_{4m_c^2}^{s_0} d s \int_{m_s^2}^{v_0} d v \int_{\alpha_{min}}^{\alpha_{max}} d \alpha \;e^{-\frac{s}{M_{1B}^2}-\frac{v}{M_{2B}^2}} (2m_c^2-3{\cal H}_\alpha) v \;,\\
\Pi^{\langle\bar{q}q\rangle}(s_0,v_0,M_{1B}^2,M_{2B}^2) &=&\frac{3 m_s \langle\bar{q}q\rangle {\cal D}}{2 \pi^2} \int_{4m_c^2}^{s_0} d s \int_{\alpha_{min}}^{\alpha_{max}} d \alpha \;e^{-\frac{s}{M_{1B}^2}}(3{\cal H}_\alpha - 2m_c^2)\;,\\
\Pi^{\langle\bar{s}s\rangle}(s_0,v_0,M_{1B}^2,M_{2B}^2) &=&\frac{3 m_s \langle\bar{s}s\rangle {\cal D}}{4 \pi^2} \int_{4m_c^2}^{s_0} d s \int_{\alpha_{min}}^{\alpha_{max}} d \alpha\; e^{-\frac{s}{M_{1B}^2}}( 2m_c^2-3{\cal H}_\alpha)\;,\\
\Pi^{\langle G^2 \rangle}(s_0,v_0,M_{1B}^2,M_{2B}^2) &=&\frac{\langle G^2 \rangle {\cal D}}{128 \pi^4} \bigg{\{}\int_{m_s^2}^{v_0} d v \int_0^1 d \alpha \; e^{-\frac{m_c^2}{\alpha(1-\alpha)M_{1B}^2}-\frac{v}{M_{2B}^2}} \bigg( \frac{3 v (m_c^2-\alpha(\alpha-1)M_{1B}^2)}{128\alpha(\alpha-1)M_{1B}^2}\nonumber\\
&+&\frac{v m_c^2[m_c^2(1-6\alpha+6\alpha^2)+3\alpha M_{1B}^2 (2\alpha^3-4\alpha^2+3\alpha-1)]}{128\alpha^3(\alpha-1)^3M_{1B}^4} \bigg)\nonumber\\
&+& \int_{4m_c^2}^{s_0} d s \int_{\alpha_{min}}^{\alpha_{max}} d \alpha 6 e^{\frac{s}{M_{1B}^2}}(2m_c^2-3{\cal H}_\alpha) \bigg{\}}\;,\\
\Pi^{\langle G^3 \rangle}(s_0,v_0,M_{1B}^2,M_{2B}^2) &=&\frac{\langle G^3 \rangle{\cal D}}{512\pi^4}\int_{m_s^2}^{v_0} d v \int_0^1 d \alpha \frac{v m_c^2 \;e^{-\frac{m_c^2}{\alpha(1-\alpha)M_{1B}^2}-\frac{v}{M_{2B}^2}}}{\alpha^4(\alpha-1)^4M_{1B}^6} [4m_c^2(2 \alpha ^4\nonumber\\
&-&4 \alpha ^3+6 \alpha ^2-4 \alpha +1) + \alpha M_{1B}^2 (4 \alpha ^5-12 \alpha ^4\nonumber\\
&+&41 \alpha ^3-62 \alpha ^2+38 \alpha -9) ]\;,
\end{eqnarray}
where $s_0$ and $v_0$ are the continuum threshold of $\eta_c$ and $K^{+\ast}$, and $M_{1B}^2$ and $M_{2B}^2$ are their Borel parameters, respectively. Taking the transformation of Eq.(\ref{three-point-p-BJ}) with $J/\psi \to K^{+\ast}$ and $K^+ \to \eta_c$, the phenomenological side of decay process $Z^+_{cs} \to\eta_c K^{+\ast}$ will be obtained.  In numerical analysis, $\sqrt{s_0} = 3.1\; \rm{GeV}$, $\sqrt{v_0} = 1.0\; \rm{GeV}$, $2.0\; \rm{GeV}^2 \le M_{1B}^2 \le 3.0\; \rm{GeV}^2$, and $2.0\; \rm{GeV}^2 \le M_{2B}^2 \le 3.0\; \rm{GeV}^2$ is the proper Borel window for decay process $Z^+_{cs} \to\eta_c K^{+\ast}$.

\subsection{The decay spectral densities of $Z^+_{cs}$ for tetraquark state}
The Ref. \cite{Dias:2013qga} discussed that only the color-connected diagrams which give the nontrivial color-structure should be considered for tetraquark states, which suggests that only the condensates of $ \langle \bar{q} G q \rangle$ and $\langle \bar{s} G s \rangle$ of OPE side need to be considered. For the color-connected diagrams, we need to isolate the $q_1^\mu q_2^\nu$ structure of both sides of Eqs.(\ref{three-point-p}) and (\ref{three-point-o}).

On the OPE side of QCD sum rules for the current Eq.(\ref{current-mixing2}), the three-point function of $Z^+_{cs} \to \bar{D}^\ast D_s^+$ after Borel transformation is
\begin{eqnarray}
\Pi^{\langle \bar{q} G q \rangle+\langle \bar{s} G s \rangle}(M_B^2)&=&\frac{m_c (\langle \bar{q} G q \rangle+\langle \bar{s} G s \rangle)}{64\pi^2}({\cal A-B+C-D}) \int_0^1\; d \alpha\; e^\frac{(2-\alpha)m_c^2}{(\alpha-1)M_B^2}\; \frac{1+\alpha}{1-\alpha}\nonumber\\
&+&\frac{m_c \langle \bar{s} G s \rangle}{32\pi^2}({\cal A-B}) \int_0^1\; d \alpha\; e^\frac{(2-\alpha)m_c^2}{(\alpha-1)M_B^2}\; \frac{1-3\alpha}{1-\alpha}
\end{eqnarray}
After Borel transformation on the Eq.(\ref{three-point-p}), the phenomenological side of three-point reads:
 \begin{eqnarray}\label{three-point-p-C}
   \Pi^{phen}(M_B^2)&=& \frac{3\lambda_{Z^+_{cs}} m_{D^\ast} f_{D^\ast} f_{D_s} m_{D_s}^2 g_{Z^+_{cs} \bar{D}^\ast D_s} }{4m_{Z^+_{cs}}^2(m_c+m_s)(m_{Z^+_{cs}}^2/4-m_{D^\ast}^2)} \; \\ \nonumber
      &\times&(e^{-m^2_{D^\ast}/M_B^2}-e^{-m^2_{Z^+_{cs}}/(4 M_B^2)})e^{-m^2_{D_s}/M_B^2}\; .
\end{eqnarray}
In our calculation $2\; \rm{GeV}^2 \le M_B^2 \le 3\; \rm{GeV}^2$ is the proper Borel window for decay process.

The OPE side of the three-point function of  the decay process $Z^+_{cs} \to \bar{D} D_s^\ast$ is just the transformation of three-point function of  the decay process $Z^+_{cs} \to \bar{D}^\ast D_s^+$ with  $\langle \bar{q} G q \rangle \leftrightarrow \langle \bar{s} G s \rangle$, and the phenomenological side of the three-point function of  $Z^+_{cs} \to \bar{D} D_s^\ast$ is just the transformation of Eq.(\ref{three-point-p-C}) 
with $D_s^+ \to \bar{D}$ and $\bar{D}^\ast \to D_s^\ast$. In numerical analysis $2\; \rm{GeV}^2 \le M_B^2 \le 3\; \rm{GeV}^2$ is the proper Borel window for decay process.

The OPE side of the three-point function of  the decay process $Z^+_{cs} \to J/\psi K^+$ is
\begin{eqnarray}
\Pi^{\langle \bar{q} G q \rangle+\langle \bar{s} G s \rangle}(M_B^2)&=&\frac{m_c (\langle \bar{q} G q \rangle+\langle \bar{s} G s \rangle)}{32\pi^2}({\cal A-B+C-D}) \int_0^1\; d \alpha\; e^\frac{m_c^2}{\alpha(\alpha-1)M_B^2}\; \frac{1}{\alpha}\;.
\end{eqnarray}
After employing Borel transformation to Eq.(\ref{three-point-p}) with $\bar{D}^\ast \to J/\psi$ and $D_s \to K^+$, the phenomenological side of three-point function of $Z^+_{cs} \to J/\psi K^+$ will be obtained.
\begin{eqnarray}\label{three-point-p-CJ}
\Pi^{phen}(M_{B}^2)&=&\frac{3 \lambda_{Z_{cs}^+}m_{J/\psi}f_{J/\psi}f_{K^+}m_{K^+}^2g_{Z_{cs}^+J/\psi K^+}}{m_s m_{Z_{cs}^+}^2 (m_{Z_{cs}^+}^2-m_{J/\psi}^2)}\nonumber\\
&\times&(e^{-m_{Z_{cs}}^2/M_{1B}^2}-e^{-m_{J/\psi}^2/M_{1B}^2})\;.
\end{eqnarray}
While the $m_{K^+}$ is very small, the contribution of the $e^{-m_{K^+}^2/M_{B}^2}$ is very close to $1$ which is neglecting in the 
Eq.(\ref{three-point-p-CJ}). In our calculation $2\; \rm{GeV}^2 \le M_B^2 \le 3\; \rm{GeV}^2$ is the proper Borel window for decay process.

The OPE side of the three-point function of  the decay process $Z^+_{cs} \to\eta_c K^{+\ast}$ writes as
\begin{eqnarray}
\Pi^{\langle \bar{q} G q \rangle+\langle \bar{s} G s \rangle}(M_B^2)&=&\frac{m_c (\langle \bar{q} G q \rangle+\langle \bar{s} G s \rangle)}{32\pi^2}({\cal A-B-C+D}) \int_0^1\; d \alpha\; e^\frac{m_c^2}{\alpha(\alpha-1)M_B^2}\; \frac{1}{1-\alpha}\;.
\end{eqnarray}
Taking the transformation of Eq.(\ref{three-point-p-CJ}) with $J/\psi \to K^{+\ast}$ and $K^+ \to \eta_c$, the phenomenological side of decay process $Z^+_{cs} \to\eta_c K^{+\ast}$ will be obtained. In our calculation $2\; \rm{GeV}^2 \le M_B^2 \le 3\; \rm{GeV}^2$ is the proper Borel window for decay process.

\end{widetext}
\end{document}